\documentclass{aa}
\pdfoutput=1 
\usepackage{appendix}
\usepackage{latexsym}
\usepackage{graphicx}
\usepackage{lscape}
\usepackage{afterpage}
\usepackage{booktabs}
\usepackage{multirow}
\usepackage{arydshln}
\usepackage{ulem}

\usepackage{ulem}
\usepackage[citecolor =blue]{hyperref}
\usepackage{amstext}

\begin{document}

\title{The peculiar emission line spectra  of core-Extremely Red  BOSS Quasars at $z\sim$2-3: orientation and/or evolution?}
\author{M. Villar Mart\'\i n\inst{1}, M. Perna\inst{1,2}, A. Humphrey\inst{3}, N. Castro Rodr\'\i guez\inst{4,5}, L. Binette\inst{6} \\
P. G. P\'erez Gonz\'alez\inst{1,6}, S. Mateos\inst{7,8}
A. Cabrera Lavers$^{4,5}$}
\institute{$^1$Centro de Astrobiolog\'ia, (CAB, CSIC--INTA), Departamento de Astrof\'\i sica, Cra. de Ajalvir Km.~4, 28850 -- Torrej\'on de Ardoz, Madrid, Spain\\  
$^2$INAF - Osservatorio Astrofisico di Arcetri, Largo Enrico Fermi 5, 50125 Firenze, Italy\\
$^3$Instituto de Astrof\'{i}sica e Ci\^encias do Espa\c{c}o, Universidade do Porto, CAUP, Rua das Estrelas, PT4150-762 Porto, Portugal \\ 
$^4$GRANTECAN, Cuesta de San Jos\'e s/n, E-38712 , Bre\~na Baja, La Palma, Spain \\
$^5$Instituto de Astrof\'\i sica de Canarias, V\'\i a L\'actea s/n, E-38200 La Laguna, Tenerife, Spain \\
$^6$Instituto de Astronom\'\i a, Universidad Nacional Autonoma de M\'exico, Apdo. Postal 70264, 04510, M\'exico D.F., M\'exico \\ 
$^7$Departamento de F\'\i sica de la Tierra y Astrof\'\i sica, Facultad de CC. F\'\i sicas, Universidad Complutense de Madrid, E-28040  Madrid, Spain  \\
$^8$Instituto de F\'\i sica de Cantabria, CSIC-UC, 39005 Santander, Spain \\
 \email{villarmm@cab.inta-csic.es}               
 }
\date{Accepted for publication in A\&A}

\abstract 
{Core-extremely red quasars (core-ERQ) have been proposed to  represent an intermediate evolutionary phase in which a heavily obscured quasar is blowing out the circumnuclear interstellar medium with very energetic outflows prior to becoming an optical quasar.}
{In this work, we investigate whether the properties of core-ERQ fit in the AGN orientation-based unification scenario.}
{We revise the  UV and optical emission line properties of core-ERQ in the context of the orientation based scenario. We use diagnostic diagrams based on UV emission line ratios
and UV-optical line kinematic information to compare the physical and kinematic gas properties of core-ERQ with those of other luminous narrow and broad line  AGN. In particular, we provide a revised  comparison of the [OIII] kinematics in 21 core-ERQ (20 from Perrotta et al. 2019 and SDSS J171420.38+414815.7, based on GTC EMIR near infrared spectroscopy) with other samples of quasars matched in luminosity with the aim of evaluating whether core-ERQ host the most extreme [OIII] outflows.}
{The UV line ratios suggest that the physical properties (for instance, density, metallicity) of the ionised gas in core-ERQ are  similar to those observed in the BLR of blue Nitrogen-loud QSOs.  The [OIII] outflow velocities of core-ERQ are, on average, consistent with those of very luminous blue QSO1, although extreme outflows are much more frequent in core-ERQ. These similarities can be explained in the context of the AGN unification model, assuming that core-ERQ are viewed with an intermediate orientation between type 2 (edge-on) and type 1 (face-on) QSOs. }
{We propose that core-ERQ are very luminous but otherwise normal quasars viewed at an intermediate orientation.  Such orientation allows a direct view of the outer part of the large BLR, from which core-ERQ UV emission originates; the extreme [OIII] outflow velocities are instead a consequence of the very high luminosity of core-ERQ.}


\titlerunning{The  emission line spectra of BOSS core-ERQ}
\authorrunning{Villar-Mart\'\i n et al.}

\maketitle

\section{Introduction}
\label{intro}

Ross et al. \citealt{Ross2015} discovered a  population of extremely red quasars (ERQ) in Data Release 10 (DR10) of the Baryon Oscillation Sky Survey (BOSS; Ross et al. \citeyear{Ross2012}, Dawson et al. \citeyear{Dawson2013}) in the Sloan Digital Sky Survey-III (SDSS-III; Eisenstein et al. \citeyear{eis11}). They  show very red colours (SDSS $r$ band to  WISE $W4$ band) similar to dust obscured galaxies (DOG). 
The authors identified 65 ERQ which 
span a redshift range of 0.28$<z<$4.36 with a bimodal distribution, with peaks at $z\sim$0.8 and $z\sim$2.5. 
Most objects are type 2 quasars (QSO2) or heavily reddened type 1 quasars (QSO1), but there is a subsample of 12 objects  presenting very  peculiar emission line to continuum properties which defy standard explanations based on extinction and/or orientation. This subclass, which was later named core-ERQ  by Hamann et al. (2017, \cite{Hamann2017} hereafter),  is the topic of this paper.

 \cite{Hamann2017} enlarged the core-ERQ catalogue to 97 objects by defining a less stringent colour condition. These have nearly uniform peculiar properties selected via $i-W3\ge$4.6 (AB system) and rest frame equivalent width of CIV$\lambda$1550 (CIV hereafter), REW$_{\rm CIV}>$100 \AA, at redshifts $z\sim$2.0-3.4.

 Core-ERQ have very high bolometric luminosities (median log$(L_{\rm bol}(\rm erg~s^{-1}))=$47.1$\pm$0.3; see \cite{Hamann2017} and \citealt{Ross2015} for a detailed characterisation and investigation of  core-ERQ properties).
They show unexpectedly flat UV spectra 
given their red UV-to-mid-IR colours and large line REW: 50\% core-ERQ have  REW$_{\rm CIV}>$150 \AA, vs. $\ll$1\% in normal blue type 1 quasars matched  in $i$ or $W3$ magnitude.
They show  signs of strong absorption in the X-rays with inferred column
densities of $N_ {\rm H}\ga$10$^{23}$ cm$^{-2}$ \citep{Goulding2018}.

Core-ERQ have peculiar, wingless rest-frame UV emission line profiles, with    full width at half maximum (FWHM) values between  between those found for very luminous type 1 and type 2  active galactic nuclei (AGN) at similar $z$. Their  median FWHM$_ {\rm CIV}$=3050$\pm$990 km s$^{-1}$ is narrower than   FWHM$_ {\rm CIV}$=5836$\pm$ 1576 km s$^{-1}$ for blue
quasars  matched  in $W3$ (\cite{Hamann2017}), but significantly broader than  FWHM$_ {\rm CIV}<$2000 km s$^{-1}$ of QSO2 \citep{Alexandroff2013} and narrow line radio galaxies (NLRG, \citealt{debreuck2000}) at similar $z$. Large blueshifts in excess of 2500 km s$^{-1}$ in CIV 
and other high-ionization  UV lines compared to the HI Balmer  and low-ionization permitted lines in the UV are also reported, which have been interpreted in terms of outflows \citep{Hamann2017}.  CIV blueshifts are common in normal QSO1 (e.g. \citealt{Gaskell1982},  \citealt{Sulentic2007}, \citealt{Runnoe2014}, \citealt{Vietri2018}). They have been widely interpreted  within the context of  accretion disk-wind models (e.g. \citealt{Richards2011})

Even more striking are the    kinematic properties of the forbidden [OIII]$\lambda\lambda$4959,5007 lines ([OIII] hereafter) revealed by  near infrared spectroscopy.
 (Zakamska et al. 2016 (\cite{zak16} hereafter), \citealt{Perrotta2019}). Core-ERQ  exhibit extremely broad and blueshifted [O III] emission, with widths ($W_{90}$) ranging between 2053 and 7227 km s$^{-1}$, and maximum outflow speeds ($V_{\rm 98}$) up to 6702 km s$^{-1}$. According to the  authors, at least 3-5 per cent of their bolometric luminosity is being
converted into the kinetic power of the observed winds, thus having  the potential of  affecting the entire host galaxy.

Core-ERQ also show peculiar line ratios.  For instance, Ly$\alpha$ is often very strongly absorbed, sometimes almost completely, resulting in very low  line intensities  relative to other lines such as CIV or NV$\lambda$1240 (hereafter, NV). They are also characterised by a high NV/CIV ratio (often $>$1.5)  and an intermediate [OIII]/H$\beta\sim$1--4 ratio (\citealt{Perrotta2019})\footnote{These values are derived using the total line fluxes, without any differentiation between NLR and BLR.} between type 1 and type 2 luminous AGN.

Disentangling the various explanations for these  puzzling core-ERQ spectral features in its own right is of great interest. Moreover, these systems might also be very relevant to studies of galaxy formation and evolution. They have been proposed to be near- or super-Eddington accreting obscured quasars, hosts of some of the most massive black holes (BH) at $z\sim$3, capable of triggering strong galactic outflows  that inhibit star formation in the early universe. They  may represent an intermediate phase in which a heavily obscured quasar is blowing out the circumnuclear interstellar medium (ISM) with very energetic outflows prior to becoming an optical quasar  (\citet{zak16},  \citet{Hamann2017}, \citealt{Goulding2018}, \citealt{Perrotta2019}).

 In the unified  model of  AGN (e.g., \citealt{Antonucci1993,Urry1995}), the orientation with respect to the observer of a dusty, obscuring central structure (torus or other) located within the BH gravitational radius of influence can explain certain differences  found between an obscured and an unobscured quasar. Such obscuring structure would block the view along some lines of sight towards the accretion disk and the clouds within the broad line region (BLR), so that these become partly or totally hidden. Independently of the specific properties of the blocking structure and the role of other factors that can influence the diversity of quasar properties, it is clear that orientation is key to explain certain differences (e.g., \citealt{Alonso2011,Ramos2011,Elitzur2012,Mateos2016}).
We investigate in this paper the role played by orientation, extinction  and the extreme luminosities of the core-ERQ on their observed emission lines and continuum spectral properties and  the possible implications on the evolutive scenario.

We adopt H$_0$ = 71 km s$^{-1}$ Mpc$^{-1}$, $\Omega_\Lambda$ = 0.73, and $\Omega_m$ = 0.27.

\section{Sample}
\label{sample}

\begin{table*}
\centering
\tiny
\begin{tabular}{llllllllllllllll}
\hline
  &   &   &  J0834+0159   &   &  &  \\  \hline
	 CIV/CIII] &  CIV/HeII & CIII]/HeII & NV/CIV & NV/HeII &  NIII]/CIII] &  NIV]/CIV   \\	
 1.7$\pm$0.3 &  $\ga$4.6  	& 	$\ga$2.3  &	1.1$\pm$0.4 &	 $\ga$3.3    & $\la$0.51  & $\la$0.25  	 \\  \hline
  Ly$\alpha$/HeII & Ly$\alpha$/CIV  &   (SiIV+OIV])/CIV  & CII/CIV &  OIII]$\lambda$1663/CIV & [OIII]/H$\beta$    \\	
 	$\ga$23  & 5.0$\pm$0.2 &  0.35$\pm$0.06 & $\la$0.14 &$\la$0.24 & 2.2  \\	\hline
  &   &   &  J1232+0912   &   &  &  \\  \hline

	 CIV/CIII] &  CIV/HeII & CIII]/HeII & NV/CIV & NV/HeII &  NIII]/CIII] &  NIV]/CIV   \\	
 5.0$\pm$0.8  &  $\ga$6.9  	& 	$\ga$1.2  &	1.7$\pm$0.1 &	 $\ga$11.7    & $\la$0.80  & $\la$0.14  	 \\  \hline
   Ly$\alpha$/HeII & Ly$\alpha$/CIV  &   (SiIIV+OIV])/CIV  & CII/CIV &  OIII]$\lambda$1663/CIV & [OIII]/H$\beta$    \\	
 $\ga$6.3	 & 0.86$\pm$0.08 &  0.59$\pm$0.07 & 0.13$\pm$0.03 &$\la$0.13 & 2.8  \\	\hline
   &   &   &  J2215-0056   &   &  &  \\  \hline
	 CIV/CIII] &  CIV/HeII & CIII]/HeII & NV/CIV & NV/HeII &  NIII]/CIII] &  NIV]/CIV   \\	
1.7$\pm$0.1 &  $\ga$4.2   &  $\ga$2.3  & 1.0$\pm$0.1   & $\ga$3.7 &  $\la$0.52 & $\la$0.29 \\  \hline
  Ly$\alpha$/HeII & Ly$\alpha$/CIV  &   (SiIV+OIV])/CIV  & CII/CIV &  OIII]$\lambda$1663/CIV & [OIII]/H$\beta$    \\	
$\ga$8.6  	& 2.1$\pm$0.3 &    0.31$\pm$0.09   &   $\la$0.29  & $\la$0.25 & 2.2 \\ \hline 

&   &   &   J2323-0100   &   &  &  \\  \hline
	 CIV/CIII] &  CIV/HeII & CIII]/HeII & NV/CIV & NV/HeII &  NIII]/CIII] &  NIV]/CIV   \\	
 4.7$\pm$0.8 &  $\ga$5.3	& $\ga$1.0	 &	1.9$\pm$0.2	&    $\ga$9.8		& 	$\la$1.1 &  $\la$0.18  \\ \hline 
  Ly$\alpha$/HeII & Ly$\alpha$/CIV  &   (SiIV+OIV])/CIV  & CII/CIV &  OIII]$\lambda$1663/CIV & [OIII]/H$\beta$   \\
$\ga$2.4  & 0.43$\pm$0.07 & 0.54$\pm$0.06 &  0.09$\pm$0.02 & $\la$0.18 & 2.1  	\\ \hline
&   &   &   J1714+4148   &   &  &  \\  \hline
	 CIV/CIII] &  CIV/HeII & CIII]/HeII & NV/CIV & NV/HeII &  NIII]/CIII] &  NIV]/CIV   \\	
  2.6$\pm$0.3 &  $\ga$3.24 	& $\ga$1.2	 &	1.8$\pm$0.2	&    $\ga$5.2		& 	$\la$0.96 &  $\la$0.38  \\ \hline 
  Ly$\alpha$/HeII & Ly$\alpha$/CIV  &   (SiIV+OIV])/CIV  & CII/CIV &  OIII]$\lambda$1663/CIV & [OIII]/H$\beta$   \\
$\ga$3.5  & 1.1$\pm$0.2 & 0.53$\pm$0.05 &  $\ga$0.35 & $\la$0.32 &  1.4$\pm$0.3  	\\ \hline
&   &   &   &   &  &  \\  
&   &   &  High $z$ NLRG &   &  &  \\  
&   &   &     &   &  &  \\   	\hline 
 CIV/CIII] &  CIV/HeII & CIII]/HeII & NV/CIV & NV/HeII &  NIII]/CIII] &  NIV]/CIV   \\	
2.5$\pm$0.1 & 1.66$\pm$0.06	& 0.64$\pm$0.03	 &	0.36$\pm$0.02 &    0.59$\pm$0.04	& 	$\la$0.1 &  0.06$\pm$0.01 \\ \hline
  Ly$\alpha$/HeII & Ly$\alpha$/CIV  &   (SiIV+OIV])/CIV  & CII/CIV &  OIII]$\lambda$1663/CIV   \\
 10.2$\pm$0.4 &  6.1$\pm$0.2 & 0.19$\pm$0.01  &  $\la$0.04 &  0.09$\pm$0.01 & 	\\ \hline

 &   &   &   &   &  &  \\  
&   &   &  High $z$ QSO2 Group 1 &   &  &  \\  
&   &   &     &   &  &  \\   	\hline 
 CIV/CIII] &  CIV/HeII & CIII]/HeII & NV/CIV & NV/HeII &  NIII]/CIII] &  NIV]/CIV   \\	
 	2.5$\pm$0.2 & 3.44$\pm$0.08 & 1.40$\pm$0.12	 &  0.41$\pm$0.03  	& 1.37$\pm$0.09	 &  N/A &  0.17$\pm$0.10 \\ \hline
  Ly$\alpha$/HeII & Ly$\alpha$/CIV  &   (SiIV+OIV])/CIV  & CII/CIV &  OIII]$\lambda$1663/CIV   \\
  18.9$\pm$3.5 &  5.5$\pm$0.9 &  0.20$\pm$0.02 &  N/A &  N/A  	\\ \hline

 &   &   &   &   &  &  \\  
&   &   &  High $z$ QSO2 Group 2 &   &  &  \\  
&   &   &     &   &  &  \\   	\hline 
 CIV/CIII] &  CIV/HeII & CIII]/HeII & NV/CIV & NV/HeII &  NIII]/CIII] &  NIV]/CIV   \\	
 	3.1$\pm$1.0  & 6.6$\pm$0.4 & 2.1$\pm$0.2	 &  0.50$\pm$0.03  	& 3.3$\pm$0.3	 &  N/A & 0.06$\pm$0.01 \\ \hline
  Ly$\alpha$/HeII & Ly$\alpha$/CIV  &   (SiIV+OIV])/CIV  & CII/CIV &  OIII]$\lambda$1663/CIV   \\
   27.6$\pm$2.0 &  4.2$\pm$0.2 &  0.18$\pm$0.02 &  N/A &  N/A  	\\ \hline

&   &   &     &   &  &  \\   
&   &   &  Radio Quiet QSO1    &   &  &  \\  
&   &   &     &   &  &  \\   \hline
 CIV/CIII] &  CIV/HeII & CIII]/HeII & NV/CIV & NV/HeII &  NIII]/CIII] &  NIV]/CIV   \\	
3.0$\pm$0.4 &   38.0$\pm$19.5  	& 	12.8$\pm$6.8 &	0.46$\pm$0.06 &  17.5$\pm$8.8   	& 0.016$\pm$0.002  & 0.016$\pm$0.003	  \\  \hline
Ly$\alpha$/HeII & Ly$\alpha$/CIV  &   (SiIV+OIV])/CIV  & CII/CIV &  OIII]$\lambda$1663/CIV   \\
100.0$\pm$50.0  & 2.6$\pm$0.3  & 0.31$\pm$0.05   &  0.026$\pm$0.011 &   0.018$\pm$0.016 \\ \hline
&   &   &     &   &  &  \\   
&   &   &  Radio Loud QSO1    &   &  &  \\  
&   &   &     &   &  &  \\   \hline
 CIV/CIII] &  CIV/HeII & CIII]/HeII & NV/CIV & NV/HeII &  NIII]/CIII] &  NIV]/CIV   \\	
3.94$\pm$0.06 &   43.7$\pm$1.8	& 	11.1$\pm$0.5 &	0.42$\pm$0.01 &  18.5$\pm$0.9   	& 0.038$\pm$0.002  & 0.054$\pm$0.002	  \\  \hline
  Ly$\alpha$/HeII & Ly$\alpha$/CIV  &   (SiIV+OIV])/CIV  & CII/CIV &  OIII]$\lambda$1663/CIV   \\
84.0$\pm$3.6  & 1.92$\pm$0.02 & 0.17$\pm$0.01  &  0.007$\pm$0.001 &  0.043$\pm$0.001  \\  \hline
&   &   &     &   &  &  \\   
&   &   & N-loud quasars  &   &  &  \\  
&   &   &     &   &  &  \\   \hline
 CIV/CIII] &  CIV/HeII & CIII]/HeII & NV/CIV & NV/HeII &  NIII]/CIII] &  NIV]/CIV   \\	
3.6  &   7.7 &  1.9  & 0.95 &  6.1 &    	 0.56	& 0.08  \\ \hline
  Ly$\alpha$/HeII & Ly$\alpha$/CIV  &   (SiIV+OIV])/CIV  & CII/CIV &  OIII]$\lambda$1663/CIV   \\
N/A & N/A &  N/A  &  N/A & 0.09 \\ \hline
&   &   &     &   &  &  \\   
\end{tabular}
  \caption{Line ratios of the four core-ERQ in \cite{zak16} and of J1744+4148.   The UV line ratios have been measured using the SDSS BOSS spectra. [OIII]/H$\beta$ taken from \citep{Perrotta2019} and our GTC-EMIR data on J1744+4148.  For comparison, UV line ratios  for three composite spectra of powerful AGN are also shown: the high $z$  narrow line radio galaxy composite ($z\sim$2.5) from Vernet et al. (\citeyear{Vernet2001}) and the radio quiet and radio loud QSO1 composites at 0.3$<z<$3.6 from Telfer et al. (\citeyear{Telfer2002}). The median line ratios of SDSS-III BOSS QSO2  at $z>$2  are also shown (Group 1, CIV/HeII$<$4 and Group 2, CIV/HeII$>$4, \citealt{Silva2019}). The bottom lines show the median values of the line ratios measured by Batra \& Baldwin (\citeyear{bat14}) in 41 N-loud quasars. Everywhere the lines are NV$\lambda$1240, HeII$\lambda$1640,  CIV$\lambda$1550, CIII]$\lambda$1909, CII$\lambda$1335, NIV]$\lambda$1486, NIII]$\lambda$1750, (SiIV+OIV])$\lambda$1400 and [OIII]$\lambda$5007. N/A means that the values are not available.}
\label{ratios}
\end{table*}

 The sample studied in this paper consists of 21 core-ERQ. We focus on the 20 core-ERQ studied by \cite{Perrotta2019} (which includes the four objects studied by \cite{zak16}), for which the authors present near infrared spectroscopy in the H$\beta$-[OIII]$\lambda\lambda$4959,5007 region. These  lines, especially the  [OIII] doublet  will provide  essential information in our argumentation.  We exclude core-ERQ-like objects (see \citealt{Hamann2017}) to have a more homogeneous sample in terms of peculiar spectral energy distributions (SED)  and emission line properties.  Detailed information on the sample can be found in \cite{Perrotta2019}. We also include the core-ERQ SDSS J171420.38+414815.7 at $z=$2.34 (J1714+4148 hereafter) that we observed with the 10.4m Gran Telescopio Canarias (GTC) (see Sect. \ref{observations}).

\subsection{Measurement of the UV line ratios}

We have measured UV line ratios for the sample of core-ERQ  using the BOSS optical spectra. As a guidance,  we show  in Table \ref{ratios} the line ratios for the four core-ERQ in  \cite{zak16} and for J1714+4148.  The UV ratios for 16 the remaining core-ERQ are shown in Appendix \ref{appendix1}. Through the text, we will refer to the  emission lines as follows: NV for  NV$\lambda$1240,  CIV for CIV$\lambda$1550, HeII for HeII$\lambda$1640, CIII] for CIII]$\lambda$1909, SiV+OIV] for SiV+OIV]$\lambda$1400.

 Ly$\alpha$ is often  blended with NV (see Fig. 18 in \citealt{Hamann2017}).
 This effect is less severe in core-ERQ than in QSO1 because  Ly$\alpha$ is often heavily absorbed and the lines are  narrower.   Isolating both lines could be done clearly in 16/21 objects. A careful evaluation of the possible contamination of NV by Ly$\alpha$ was necessary in five objects with severe blending (see, for instance, spectra of J0834+0159 and J2215-0056 in Fig. 1 of \cite{zak16}).

To estimate this effect, we proceeded as follows. Maximum contamination might be expected  for the highest possible Ly$\alpha$  flux. This is, for no Ly$\alpha$ absorption and maximum flux relative to NV. As a reference to estimate this, we have used   the measured  CIV   fluxes   and the  typical  Ly$\alpha$/CIV ratios observed  in AGN.  

A  range of values Ly$\alpha$/CIV$\sim$1-20 is measured in luminous type 2 AGN  (e.g. \citealt{debreuck2000}, \citealt{Villar2007}, \citealt{Alexandroff2013}). The lowest values are a consequence of prominent Ly$\alpha$ absorption. The ratio is typically  Ly$\alpha$/CIV$\la$6  in QSO1 (\citealt{Zheng1997}, \citealt{Lusso2015}, \citealt{Telfer2002}). 

For each object we have created artificial Ly$\alpha$ spectral profiles for which  Ly$\alpha$/CIV=20.  The Ly$\alpha$ FWHM and velocity offset relative to NV are set to be equal to those of the strongest emission lines (CIV, SiIV).  This "unabsorbed" Ly$\alpha$ profile would result in a maximum possible contamination of NV. The strongest emission lines in our sample have spectral profiles reasonably well fitted with a Gaussian. Consistently, in our analysis we have assumed  a  Gaussian shape for the  unabsorbed Ly$\alpha$. Given that the UV lines of each target typically show a range in their FWHM and in their  redshifts $z_{i}$, we have assumed the worst case scenario and adopted as Ly$\alpha$ FWHM the broadest value observed among the different lines.  It was found that uncertainties on the relative velocity ($z_{\rm Ly\alpha}$) with respect to NV did not have a significant impact on the results.

The  uncertainties of the NV flux derived from the above evaluation were  $\la$20\% for four of the  five objects under consideration and $\sim$30\% for a fifth object.  Since Ly$\alpha$ is heavily absorbed in general and the worst case scenario has been adopted when estimating the Ly$\alpha$ contamination, we can reasonably conclude that its impact is in general low. The errors hereafter quoted about the line ratios involving NV will include all the above mentioned uncertainties. These can therefore be qualified as the maximum errors expected.

\begin{figure*}
\centering
\includegraphics[width=1.0\textwidth,height=0.40\textwidth]{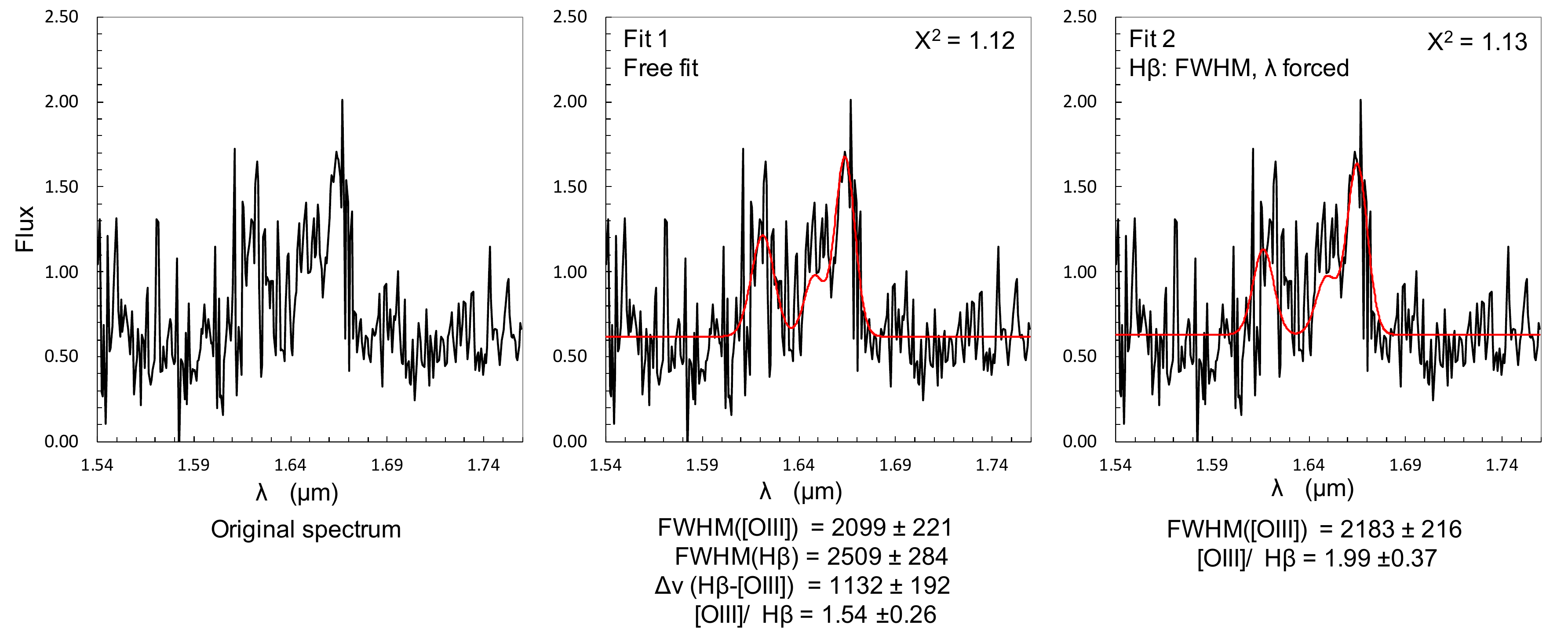}
\includegraphics[width=1.0\textwidth,height=0.40\textwidth]{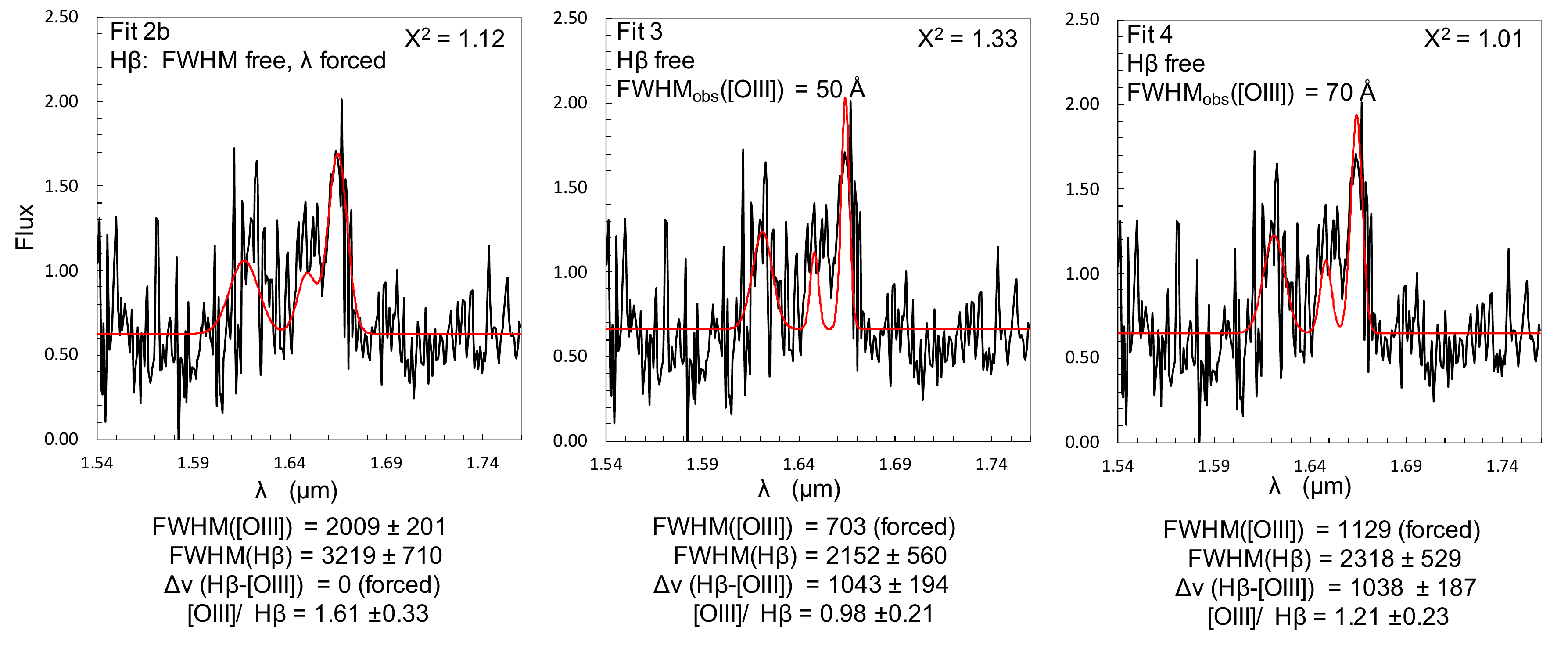}
\includegraphics[width=1.0\textwidth,height=0.40\textwidth]{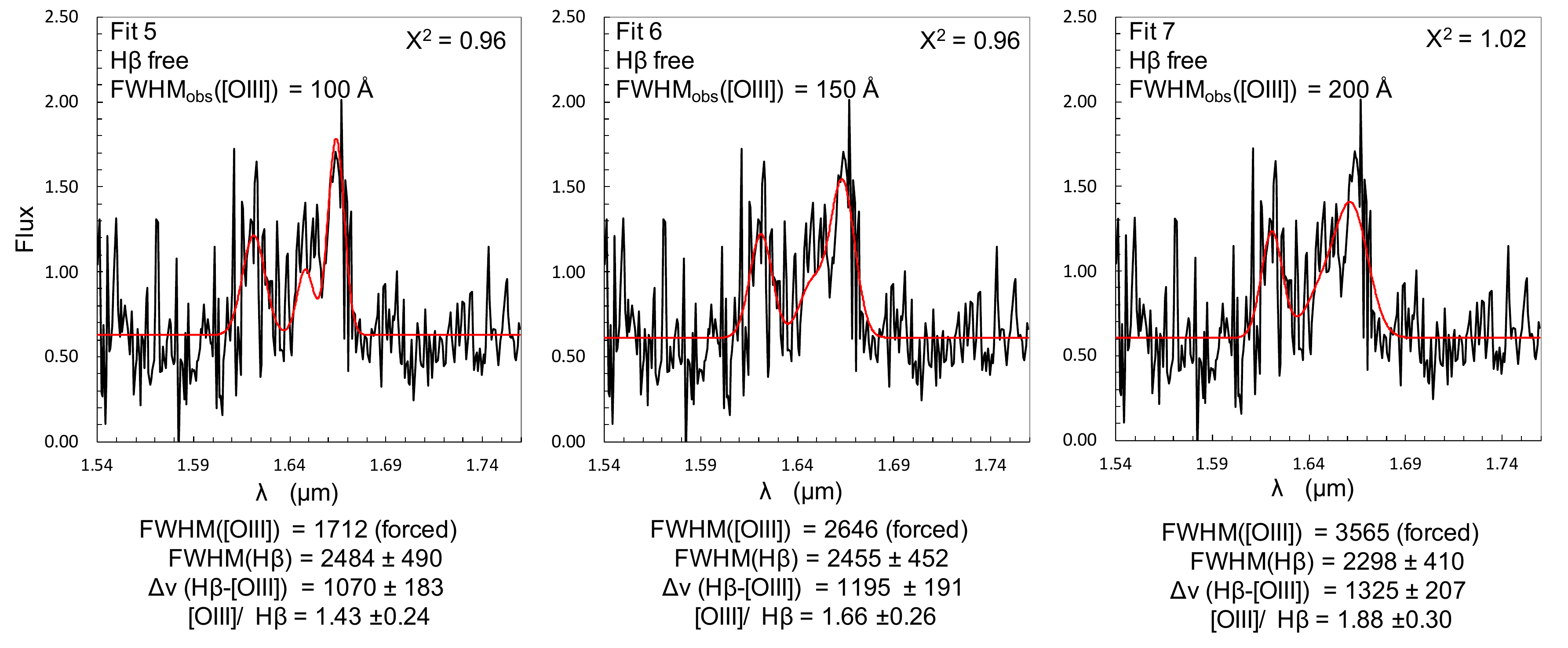}
\vskip2mm
\caption{GTC EMIR spectrum of J1714+4148 in the H$\beta$ and [OIII]  spectral region (black  lines). The flux is in units of $\times$10$^{-18}$ erg s$^{-1}$ cm$^{-2}$  \AA$^{-1}$.  Different fits (red lines) were attempted by applying a variety of kinematic constraints to [OIII]  and/or H$\beta$ (fits 2 to 7) or by leaving the kinematic parameters free (fit 1). For each fit, the constraints on the observed FWHM in \AA~ are shown. It is also specified whether H$\beta$ was forced to have the same redshift as [OIII]$\lambda$5007  ($\lambda$ forced, fits 2 and 2b). The reduced chi-squared $\chi^2$ are also shown, and the best-fit profiles are presented underneath each panel.}
\label{figemir}
\end{figure*}

\section{J1714+4148: GTC EMIR observations, data
reduction and analysis}
\label{observations}

We obtained HK-band spectroscopy of the core-ERQ  J1714+4148 at $z=$2.34 with the 10.4m Gran Telescopio Canarias (GTC) and the EMIR (Espectr\'ografo  Multiobjeto Infra-Rojo)  instrument in long slit mode (program GTC14-19A). EMIR is a near-infrared 
wide field imager and medium-resolution multi-object spectrograph installed at the Naysmith-A focal station. It is equipped with a 2048$\times$2048 Teledyne HAWAII-2 HgCdTe NIR-optimised chip with a pixel size of 0.2$\arcsec$. The low resolution HK  grism covers a spectral range of $\sim$1.45-2.42 $\mu$m  with a dispersion of 6.8 \AA\  pixel$^{-1}$. The slit width used during the observations was 1.0$\arcsec$. The instrumental profile measured from the OH sky lines is FWHM$_{\rm IP}$=31.3$\pm$4.0 \AA.  The total exposure time on source was 3840 seconds.  A typical ABBA nodding pattern was applied. The seeing during the observations was  FWHM$\sim$0.7$\arcsec$. 

The spectra were reduced using several python routines customised by GTC staff for EMIR spectroscopic data. The sky background was first eliminated   using consecutive A-B pairs. They were subsequently  flat-fielded, calibrated in wavelength and combined to obtain the final spectrum. 

To correct for telluric absorption, we observed a telluric standard star with the same observing set up as the science target, right  after the J1714+4148 observations and at similar airmass. To apply the correction we used a version of Xtellcor (\citealt{Vacca2003}) specifically modified to account  for the atmospheric conditions of  La Palma observatory (\citealt{Ramos2009}). 

 Relative flux calibration was applied using the spectrum of the  star, which was obtained with the same narrow 1.0$\arcsec$ slit. The accuracy is $\sim$10\%. It is not clear how different this is from the absolute flux calibration, since  near infrared magnitudes are not available for the object. Based on previous experience, we estimate a maximum deviation of $\la$30\% between the relative and absolute flux calibrations.   Since galactic extinction is very low ($A_{\rm V}$=0.06) correction for this effect was not applied.

We fitted H$\beta$ and [OIII]$\lambda\lambda$4959,5007 with single Gaussian profiles to obtain an approximate quantitative characterisation of the gas kinematics.  The separation in $\lambda$ and flux ratio (3:1) for the [OIII] doublet were  fixed to the theoretical values. Each line is parametrised with a central $\lambda$, observed FWHM and  amplitude (and thus, the flux). The two [OIII] lines were forced to have the same FWHM.  More complex profiles are likely to be more realistic (\citealt{Perrotta2019}) but this approach is not possible with our data given the low S/N  ([OIII]$\lambda$5007 is detected with S/N$\sim$6).  All FWHM values in km s$^{-1}$ quoted below have been corrected for instrumental broadening in quadature.

We show in Fig. \ref{figemir} (first panel) the H$\beta$ and [OIII] spectrum  and the best-fits for different sets of constraints. We attempted different fits applying  a variety of kinematic constraints to determine  useful ranges of the lines FWHM and the shift in velocity  $\Delta v_{\rm H\beta-[OIII]}$.  Some examples  are shown in  Fig. \ref{figemir}.  Each panel  shows the fit with the smallest  mathematical errors (minimum reduced chi-squared,  $\chi^2$) for that specific set of constraints.  

The low S/N of the spectrum prevents accurate constraints on the [OIII] and H$\beta$ parameters. In fact, the visual comparison between the data and the fits shows  that $\chi^2\sim$1 is not always associated with satisfactory results (see, for instance, fits 4 and 7).

 From this analysis, we propose two tentative results: the [OIII] lines are broad, with a most probable range 1700$\la$ FWHM$_{\rm [OIII]}<$2300 km s$^{-1}$ (100$\la$ FWHM$_{\rm [OIII]}^{\rm obs}<$130\AA. The fits suggest that  H$\beta$ is broader than [OIII] and is redshifted, although we cannot discard that this is an artificial result due to the noise of the H$\beta$ profile. Tentatively, 2000$\la$ FWHM$_{\rm H\beta}\la$3000 km s$^{-1}$ and  900$\la\Delta v _ {\rm H\beta-[OIII]}\la$1300 km s$^{-1}$. Such broad H$\beta$ and [OIII] lines and prominent [OIII] blueshifts  are common in core-ERQ (\citealt{Perrotta2019}).

\section{Results and discussion}
\label{results}

\subsection{Comparison of the UV  emission line ratios with blue QSO1, QSO2 and high $z$ NLRG}
\label{compar}

\begin{figure*}
\hskip1cm
\includegraphics[width=0.9\textwidth,height=0.45\textwidth]{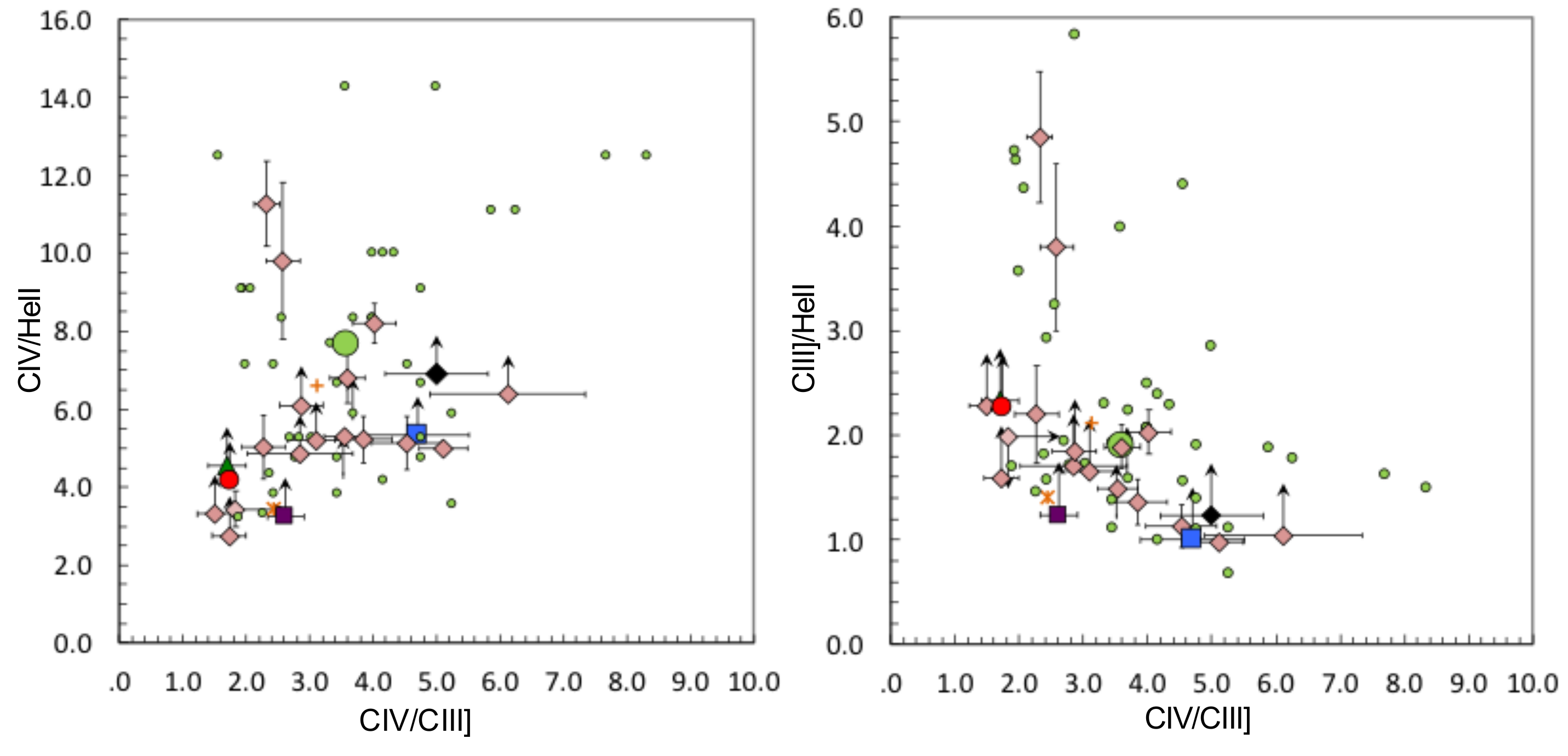}
\vskip2mm
\hskip1cm
\includegraphics[width=0.9\textwidth,height=0.45\textwidth]{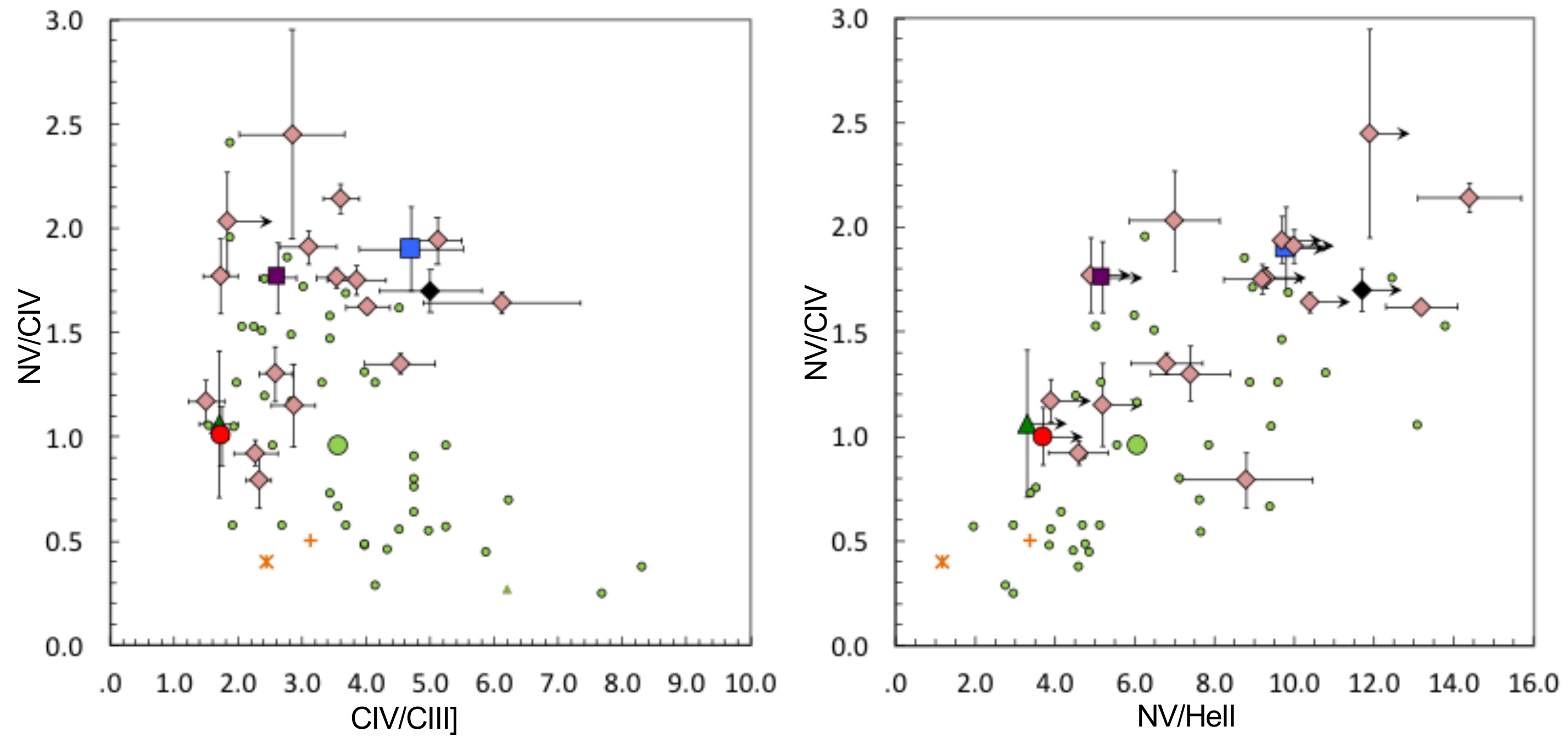}
\vskip2mm
\hskip7cm
\includegraphics[width=0.3\textwidth,height=0.1\textwidth]{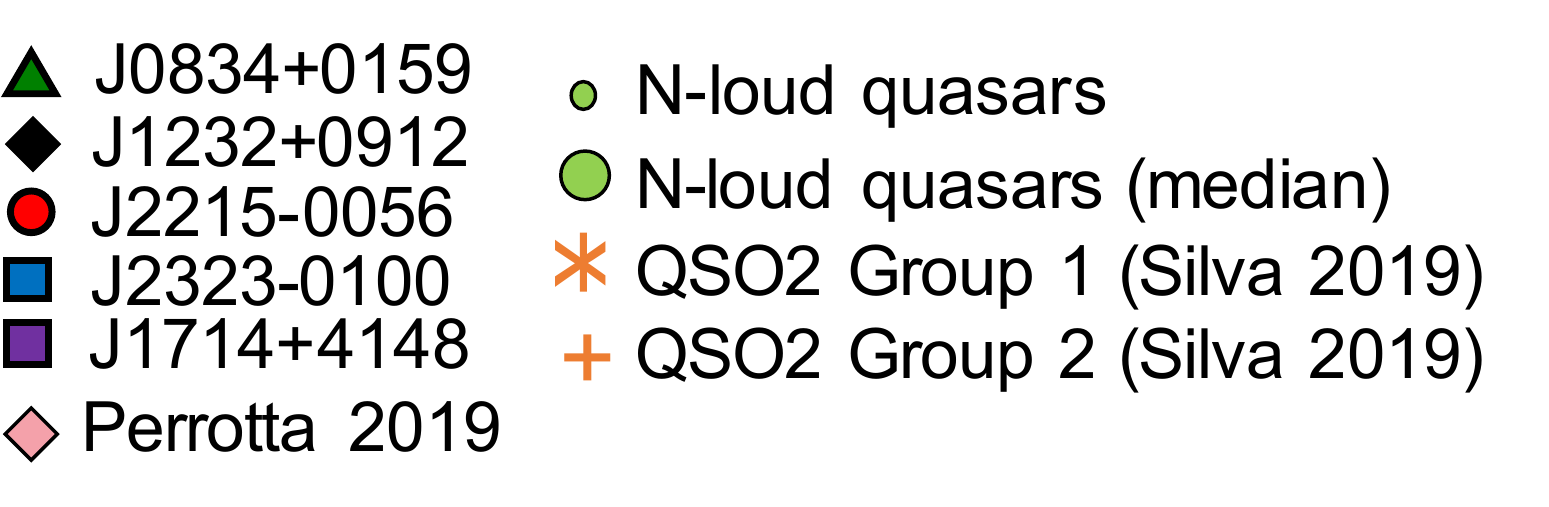}
\caption{Comparison of the core-ERQ UV line ratios with nitrogen-loud (N-loud) quasars  (\citealt{bat14}) and the median ratios of SDSS QSO2 at $z>2$ (Group 1: CIV/HeII$<$4 and Group 2: CIV/HeII$>$4, \citealt{Silva2019}). Since HeII$\lambda$1640 is not detected for most core-ERQ,  lower limits are shown with arrows for the ratios involving this line.   The minimum and maximum y-axis and x-axis values have been selected in all diagrams to span approximately the full range of values  shown by most  objects plotted. The core-ERQ line ratios (measured and lower/upper limits)  overlap partially with  the area of the diagrams covered by N-loud quasars.}             
\label{brotherton1}
\end{figure*}

 The line ratios of the four  core-ERQ in \cite{zak16} and J1714+4148  are shown in  Table \ref{ratios} (see  Appendix \ref{appendix1}  for  the  remaining 16 objects in the sample). The UV line ratios of three composite spectra of other powerful AGN  samples are also shown for comparison: NLRG, radio loud (RL) and radio quiet (RQ) QSO1 (\citealt{Telfer2002}). The  NLRG composite  (\citealt{Vernet2001}) was created by combining high signal to noise Keck spectra of 9 powerful radio galaxies at $z\sim$2.5.  The RQ and RL  QSO1 composites (\citealt{Telfer2002}) were created by combining Hubble Space Telescope (HST)  spectra of QSO1 at 0.3$\la$z$\la$3.6. Most objects (139 out of 184) are at $z\la$1.5. The RQ and RL  composites include spectra of 107 and 77 objects respectively.  The  median line ratios of the sample of nitrogen-loud (N-loud) QSO1 from \cite{bat14} and the median line ratios of SDSS-III BOSS QSO2 candidates at 2$\la z\la$4.3 (\citealt{Alexandroff2018}, \citealt{Silva2019}) are also shown. 
The location of the 21 core-ERQ in several diagnostic diagrams involving relevant UV lines is shown in Fig. \ref{brotherton1}.

\subsubsection{Comparison with luminous type 2 AGN}

The comparison with NLRG  is interesting because these are obscured AGN with quasar like luminosities (\citealt{Vernet2001}). The BLR is hidden and the emission line spectrum originates in the NLR.    As shown by \cite{humphrey2008}, the  UV and optical emission line ratios of high $z$ NLRG are best explained by AGN photoionization   of low density gas ($n < $1000 cm$^{-3}$). 

The  line ratios of the  core-ERQ  are inconsistent with  NLRG (Table \ref{ratios}, see also \citealt{Silva2019}). NLRG show  much stronger HeII relative to other lines and  fainter NV. Typically,  CIV/HeII$<$2,  CIII]/HeII$<$1, NV/CIV$\la$1 and NV/HeII$\la$1 in NLRG  (Villar Mart\'\i n et al. \citealt{vm97}, \citealt{debreuck2000}, \citealt{humphrey2008}; 
see also Table \ref{ratios}). As can be seen in Fig.   \ref{brotherton1},   in general, core-ERQ show CIV/HeII$\ga$3, CIII]/HeII$\ga$1, NV/CIV$\ga$1 and NV/HeII$\ga$3.
 Star forming galaxies also show relatively weak HeII, with  high CIV/HeII and CIII]/HeII  similar to core-ERQ. However,  stellar photoionization can be discarded with confidence as  a relevant excitation mechanism  of the gas on the basis of the huge EW of the emission lines and the strength of NV  (e.g. \citealt{Feltre2016}, \citealt{Nakajima2018}).

\cite{Alexandroff2013} identified a sample of 145 QSO2 candidates at 2$\la z\la$4.3  selected from the quasar sample of the SDSS-III BOSS. They  have weak rest-frame UV continuum (typical continuum magnitude of $i\sim$22) and strong CIV and Ly$\alpha$, with FWHM$\la$2000 km s$^{-1}$. \cite{Silva2019} have recently studied the physical properties and abundances of the ionized gas in these systems based on their location in rest-frame UV diagnostic diagrams  and on the comparison  with   AGN photoionization model predictions. Compared to NLRG  at similar redshifts, the QSO2 are offset to higher NV/HeII, CIV/HeII and CIII]/HeII. 

 \cite{Silva2019} have classified the QSO2 in two groups: objects with `normal' CIV/HeII ratios of $<$4 (Group 1), and those with 'extreme' CIV/HeII ratios of $>$4  (Group 2). Group 2 QSO2 also have systemically higher NV/HeII ratios for otherwise similar line ratios. To explain such systems as well as the difference with Group 1 QSO2 and high z NLRG, \cite{Silva2019} propose a combination of high gas density $n\ga$10$^7$ cm$^{-3}$ and/or supersolar abundances $Z\ga$4$\times Z_{\rm \odot}$ with 
N/H$\ga$16$\times Z_{\rm \odot}$ assuming secondary production of N that results in a quadratic (rather than linear) increase of its N/H abundance. The secondary production of C is also discussed by these authors as a possible explanation for the extreme CIV/HeII ratios seen in some QSO2s. However, in the case of core-ERQ no models can explain the line ratios simultaneously.

 Based on the value CIV/HeII$>$4,  17 of the 21 core-ERQ  studied here are confirmed to belong to Group 2 (Fig. \ref{brotherton1}, tope left panel). The classification of the remaining 4 is not possible based on the CIV/HeII lower limits. However,  the high NV/HeII (lower limits in the range  $\sim$4-7) strongly suggest that they also belong to Group 2 (NV/HeII$<$3 for Group 1 QSO2, \citealt{Silva2019}). The same classification appears to apply to core-ERQ in general, based on the visual inspection of their spectra  (\citealt{Hamann2017}).

The four core-ERQ studied by \cite{zak16}  roughly show the lower (J0834+0159 and J2215-0056)   and upper values J1232+0912 and J2323-0100) of the range of line ratios  spanned by core-ERQ (Fig. \ref{brotherton1}; let us not forget that only upper limits could be measured for HeII). We
 highlight them with different symbols for guidance  and as rough markers of   the extreme  locations in the diagrams of  core-ERQ.   Most of the remaining core-ERQ  overlap with them or lie in between.

We  discuss next the optimum AGN photoionization models within the range explored by \cite{Silva2019}. They consider $n$ in the range 10$^{2-8}$ cm$^{-3}$ and gas metallicity in the range $Z=$ (0.5-5.0)$\times Z_{\rm \odot}$.

$\bullet$ J0834+0159 and J2215-0056.  They have among the lowest UV line ratios of core-ERQ. Based on their location in the diagrams NV/CIV vs. NV/HeII, NV/CIV vs. CIV/CIII], CIV/HeII vs. CIV/CIII],   the optimum models have the following parameters (see Fig. 1 in \citealt{Silva2019}):

\begin{itemize}

\item   Secondary   N and primary C models. Models with  $n\sim$10$^{6-8}$ cm$^{-3}$  and 3-5 $Z_{\odot}$ are consistent with the location of the two objects in all  diagrams mentioned above. These models are also consistent with NIV]/CIV$\la$0.3 measured in these two objects. 

\item No adequate secondary N and secondary C models are found.  Only models with $n\sim$10$^{2-4}$ cm$^{-3}$ and $Z\sim$2$Z_{\rm \odot}$   can reach NV/CIV$\sim$2.
However, these fail in other diagrams. For instance, they produce NV/HeII$\sim$1  (compared with the observed NV/HeII$>$3). Also, they produce CIV/CIII]$\ga$5, compared with the measured CIV/CIII]=1.7.

\end{itemize}

Therefore,  within the range of models explored by \cite{Silva2019},  those favoured for J0834+0159 and J2215-0056 have primary production of C and secondary production of N, $n\sim$10$^{6-8}$ cm$^{-3}$  and $Z\sim$3-5$Z_{\odot}$.

$\bullet$  J1232+0912 and J2323-0100. Their line ratios  are rather extreme. NV is very strong compared with QSO2: NV/CIV$\sim$2 and NV/HeII$\ga$10, while QSO2 show in general NV/CIV$<$1 and NV/HeII$<$10. CIV/CIII]$\sim$5 is also at the high end of the range of values spanned by QSO2. There are no optimum models within the parameter space explored by \cite{Silva2019} that can explain all line ratios simultaneously. On the other hand, the high NV ratios and large CIV/HeII strongly suggest high densities $n>$10$^6$ cm$^{-3}$ and/or high metallicities $Z\ga$4$Z_ {\rm \odot}$.

The discussion above suggests that  densities $n\ga$10$^6$ cm$^{-3}$ and well above solar metallicities are unavoidable in the gas responsible for emitting the UV lines in  the four core-ERQ.  Since these ratios span roughly the most extreme values among core-ERQ line ratios  (Fig. \ref{brotherton1}), we extend this conclusion to core-ERQ in general. 
We will next argue  that the UV emission lines in core-ERQ have a significant, even dominant contribution from the BLR  (see also \citealt{Alexandroff2018}).

\subsubsection{Comparison with QSO1}

The relative weakness of HeII is characteristic of QSO1. The UV emission lines in these systems are preferentially emitted in the BLR. They show   typical CIV/HeII$>$35,  CIII]/HeII$>$10, NV/HeII$>$15  (Table \ref{ratios}; see also \citealt{Vandenberk2001}).   Moreover, most core-ERQ  show enhanced NV  in comparison with  the QSO1 composites (NV/CIV$\sim$0.45; see also \citealt{Vandenberk2001}).  The 97  objects in \cite{Hamann2017}  have median NV/CIV=1.44 and 90\%  show  NV/CIV$>$0.5.  65\%  show NV/CIV$>$1.0  in the range $\sim$1.0-2.9.

The enhanced NV emission  is reminiscent of N-loud quasars. The strong UV Nitrogen lines in these systems have been interpreted in terms of supersolar metallicities in the BLR (e.g. \citealt{Hamann1993}, \citealt{Dietrich2003},    \citealt{Baldwin2003}, \citealt{Nagao2006}, \citealt{Matsuoka2011}), sometimes as extreme as $Z>$10$Z_{\rm \odot}$ (e.g. \citealt{Baldwin2003}).  
 \citet{bat14} analysed the UV spectra of 41 N-loud quasars at $z\sim$2.0-3.5. They show median NV/CIV=0.95 and NV/CIV$\sim$1.0-2.4 for half the objects.  The  inferred metallicities   are in the range $\sim$1-18 $Z_{\rm \odot}$ with median  5.5 $Z_{\rm \odot}$.

\begin{figure}
\centering
\hskip-2.5mm
\includegraphics[width=0.50\textwidth,height=0.49\textwidth]{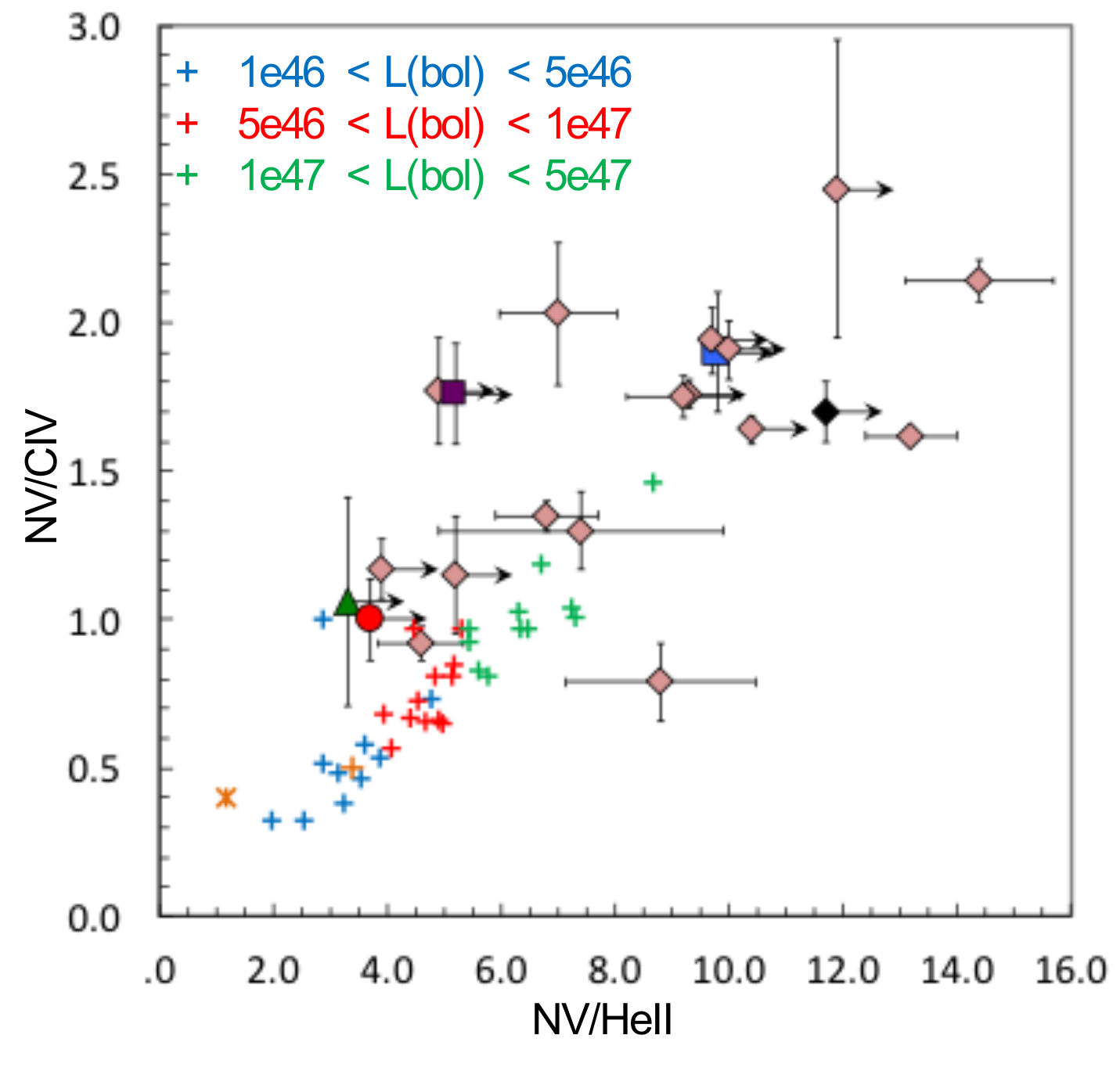}
\caption{NV ratios and BLR metallicity. The blue, red and green symbols correspond to the QSO1 bins in which \cite{Matsuoka2011} distributed 2383 SDSS QSO1 at 2.3 $< z <$ 3.0 based on the $M_{\rm BH}$ and the  Eddington ratio, $\frac{L_{\rm bol}}{L_{\rm Edd}}$.   Each cross represents the location of one given  $L_ {\rm bol}$ bin. The colours indicate ranges of increasing $L_{\rm bol}$. The median values of  QSO2 (Groups 1 and 2) are also shown for comparison with orange symbols (see text).}          
\label{matsuoka}
\end{figure}

The  UV line ratios of \citet{bat14} N-loud QSO1 are shown in the diagnostic diagrams of Fig.  \ref{brotherton1} .  
The two \cite{Telfer2002} QSO1 composites  do not appear  because their large ratios relative to HeII place them outside three of the diagrams.   It can be seen that core-ERQ   overlap  partially in the diagrams with the area covered by N-loud QSO1.   This relation is likely to apply to the general population of core-ERQ, given the similar line ratios suggested by the visual inspection of their spectra (\cite{Hamann2017}).

This  suggests that the UV lines in  core-ERQ  are emitted in metal rich BLR.

Different authors have reported a correlation between quasar $L_ {\rm bol}$ and the BLR metallicity $Z$,  as traced by the Nitrogen (N$^{+2}$, N$^{+3}$, N$^{+4}$) emission line ratios (\citealt{Hamann1999}, \citealt{Warner2004}, \citealt{Shemmer2004},  \citealt{Nagao2006}, \citealt{Juarez2009}). Given the high $L_{\rm bol}$ of core-ERQ, high BLR abundances (and the enhacement of the NV emission as a consequence) are similarly expected if they follow the same correlation.  According to \cite{Xu2018},  $Z/Z_{\rm \odot}\sim$7-16$Z_{\rm \odot}$ are expected in the BLR of quasars with log$(L_{\rm bol})\sim$47.0-48.0 (i.e., similar to core-ERQ).

To investigate the origin of the $L_{\rm bol}$-$Z_{\rm BLR}$ relation, \cite{Matsuoka2011}  used the optical
spectra of 2383 quasars at 2.3 $< z <$ 3.0 extracted from SDSS. They divided the sample into bins based on $M_{\rm BH}$ and the  Eddington ratio $\frac{L_{\rm bol}}{L_{\rm Edd}}$. They then extracted composite spectra for each bin of $M_ {\rm BH}$ and $\frac{L_{\rm bol}}{L_{\rm Edd}}$ and proceeded to measure the line ratios most sensitive to BLR metallicity, which include NV/CIV and NV/HeII.  They found that the most lumoninous QSO1 show the highest NV ratios (see Fig.  \ref{matsuoka}).  We investigate next  how NV/CIV and NV/HeII for core-ERQ behave in relation to this QSO1 correlation. 

Because no reliable $M_ {\rm BH}$ are available for core-ERQ (see Section \ref{orientation}), we investigate the behaviour of the line ratios with $L_{\rm bol}$ instead. $L_{\rm bol}$   for the different bins in   \cite{Matsuoka2011} QSO1 sample have been obtained using $\frac{L_{\rm bol}}{L_{\rm Edd}}$ and $M_{\rm BH}$ in their Table 3. They are in the range 46.34$\le$log($L_{\rm bol}$)$\le$47.54. For comparison,  46.98$\le$log($L_{\rm bol}$)$\le$47.86 for the core-ERQ discussed in this work (\cite{Perrotta2019}). 
 We plot  the QSO1 bins and  the core-ERQ ratios in the NV/CIV vs. NV/HeII diagram (Fig. \ref{matsuoka}).   QSO1 are represented with blue, red and green colours in order of  increasing $L_{\rm bol}$.  
The tight correlation between NV/CIV and NV/HeII for QSO1 reflects at least partially the lack of Baldwin effect of NV$\lambda$1240 in type 1 AGN at $z\la$5 (\citealt{Dietrich2002}). While the optical and UV broad and narrow lines of type 1 AGN show an anticorrelation between the REW  and  the continuum luminosity $\lambda L_{1450 \AA}$ (the so called   Baldwin effect, \citealt{Baldwin1977}), the REW of NV$\lambda$1240  remains nearly constant over 6 orders of magnitude in continuum luminosity (\citealt{Dietrich2002}). The different behaviour of this line has been explained as a consequence of the increasing gas metallicity with AGN luminosity   and the secondary production of N; this is, N/O $\propto$ O/H (\citealt{Korista1998}). 

Core-ERQ  show a large scatter, in part due to the non-detection of HeII in most of them. In spite of this, it appears clear that the NV ratios in these systems are among the highest compared to \cite{Matsuoka2011} QSO1, as we would expect based on their  extreme $L_ {\rm bol}$. This supports that the lines are emitted in metal-rich BLR. 

The above comparison should be taken with caution. NV  ratios are  very sensitive to the ionization level of the gas (\citealt{humphrey2008}, \citealt{bat14}). To isolate ionization effects from those due to the gas abundances, it would be  necessary to measure other lines such as NIV] and NIII] (\citealt{bat14}), which are in general undetected in the SDSS spectra of the core-ERQ. In spite of this limitation, the comparison presented above with QSO1 strongly suggest that the UV lines in core-ERQ are emitted by metal rich BLR (see also \cite{Hamann2017} and \citealt{Alexandroff2018}).

\subsection{The size  and density of the UV  emission line region}
\label{regions}

We argued above that the UV lines are emitted in the   BLR of core-ERQ. The supression of the UV/optical continuum, the wingless  line profiles and the intermediate FWHM between QSO1 and QSO2 (see Sect. \ref{orientation}) suggest that we are observing the outskirts of this region.

The range in density in the BLR  comes mainly from the estimated radii and photoionization theory (e.g. \citealt{Osterbrock1989}, \citealt{Ferland1992}). Clouds with densities from $\sim$10$^9$ cm$^{-3}$ (close to  $n_{\rm crit}\sim$3.2$\times$10$^9$ cm$^{-3}$ of C III]$\lambda$1909)  to $>10^{13}$ cm$^{-3}$ are expected.  The highest density clouds $n>$10$^{13}$ cm$^{-3}$ are continuum radiators \citep{Rees1989}.    Thus, we assume $n\ga$10$^9$ cm$^{-3}$.

The BLR in AGN is photoionized. The main evidence  is that the emission-line spectra change in response to changes in the continuum, with lag-times
corresponding to characteristic radii of the BLR \citep{Peterson1993}. Assuming, thus, photoionization, the radiation pressure likely
confines the ionized layer of the illuminated gas    \citep{ste14}. An implication is that the gas density near the ionization front varies with distance from the nucleus $r$ as:
 $$n \sim 7 \times 10^4 ~L_{\rm i,45} ~r_{\rm 50}^{-2} ~~\rm{cm^{-3}~~[eq.1]}$$

  where $L_{\rm i,45}$ is the ionizing luminosity $L_{\rm i}$ in units of 10$^{45}$ erg s$^{-1}$ and $r_{\rm 50}$ is $r$ in  units of 50 pc.   Knowing $n$ and $L_{\rm i,45}$  we can thus infer  distances.

 We have estimated $L_{\rm i}\sim 0.35 \times L_{\rm bol}$ \citep{ste14}.  If the UV lines we  see are preferentially  emitted in  a region of  $n\sim$10$^9$ cm$^{-3}$,  $r_ {\rm UV}\sim$2.1-6.7 pc or 0.7-2.1 pc for  $n\sim$10$^{10}$ cm$^{-3}$, depending on the object in the sample of 20 core-ERQ in \cite{Perrotta2019} (J1714+4148  is not considered here because $L_{\rm bol}$ is not available).

The radial size of the BLR, $r_{\rm BLR}$ of $z\la$0.3 type 1 AGN with $log(\lambda L_{\rm 5100})\la$45.0 correlates with the rest-frame 5100 \AA\  luminosity $\lambda L_{\lambda \rm 5100}$ as  $ r_{\rm BLR} = (22.3 \pm 2.1) \times \bigg(\frac{\lambda L_{\lambda \rm 5100}}{10^{44}}\bigg)^{0.69\pm0.05} {\rm lt-days}$ (\citealt{Kaspi2005}). This relation  was expanded up to $log(\lambda L_{\rm 5100})\sim$46.0 by  \cite{Vestergaard2002}.  We constrain next $r_{\rm BLR}$ for the core-ERQ, assuming that they follow this correlation.

We have inferred $\lambda L_{\lambda \rm 5100}$   following $log(L_{\rm bol}) = (4.89\pm 1.66) + (0.91\pm0.04)  \times log(\lambda L_{\rm 5100})$ \citep{Runnoe2012}.  
If  core-ERQ  follow this correlation, we expect log($\lambda L_{\lambda \rm 5100})\sim$46.1-47.2 for the 20 core-ERQ. These  are $\sim$7-45 times higher than the values inferred from the observations \citep{Perrotta2019}. This shows that  the optical continuum   at 5100 \AA~ is strongly suppressed (\citealt{Hamann2017},\citealt{Perrotta2019}), as naturally expected in our proposed scenario.
The  $\lambda L_{\lambda \rm 5100}$ inferred from $L_ {\rm bol}$ imply $r_{\rm BLR}\sim$0.6-3.1 pc depending on the object,  which are in reasonable agreement with the $r_{\rm UV}$ values estimated for $n\sim$10$^{9-10}$ cm$^{-3}$.  Because of their high luminosity, the  expected BLR sizes are  $\sim$24-126 times larger than the median size inferred for the low $z$ type 1 AGN sample  (29.2 lt-days, \citealt{Kaspi2005}).  The sizes could be somewhat larger if  obscuration makes any of the near-and mid-infrared emission anisotropic, since the intrinsic luminosities would be even higher (\cite{Hamann2017}). Core-ERQ are therefore expected to have large BLR due to their extreme luminosities.  This may favour the  partial visibility at a wider range of inclinations than for less luminous quasars.

\subsection{Orientation and partial view of the broad line region}
\label{orientation}

The impact of orientation on the properties of the SED and the emission line spectra of broad line active galaxies (for instance, quasars and broad line radio galaxies, BLRG) has been widely investigated in the literature (e.g. \citealt{Marziani2017}, \citealt{Liu2018}). Some authors have proposed  that at least a fraction of BLRG are  partially obscured, misdirected  quasars, seen at somewhat larger angles than the quasar population. In this scenario, they are  quasars seen through the edge of the obscuring torus (\citealt{Dennett2000}, \citealt{Morganti1997})  which, thus,  does not have a distinct edge (\citealt{Baker1997}). 

The role of orientation in the context of core-ERQ has also been discussed (\citealt{Hamann2017}).  \cite{Alexandroff2018} highlight its essential role in the model they propose to explain the UV  emission line and continuum spectropolarimetric properties of the core-ERQ SDSS J1652+1728 at $z$=2.94. The authors describe  a polar outflow model seen at an intermediate orientation  between  type 1 (face on) and  type 2 (edge-on) with the  UV lines being produced   on spatial scales similar or greater than the scales of the dusty broad emission and broad absorption emission line regions (see also \citealt{Hamann2017}). This scenario is in turn  inspired on the model proposed by \cite{Veilleux2016} for the nearest quasar MRK231.

Next we review and reinforce  with new arguments the significant role that orientation plays  in determining the observed properties of core-ERQ. Based on all arguments combined, we propose  that a significant fraction of core-ERQ are very luminous  but otherwise normal quasars seen at an intermediate orientation between type 1 (face-on) and type 2 (edge-on). This results  in a partial view of the BLR, with the outskirts being visible, while the inner BLR and the AGN continuum remain hidden.

This scenario is consistent with: 

$\bullet$ The suppression of the UV  and optical continuum compared with blue QSO1 (Sect. \ref{regions}, \citealt{Hamann2017})

$\bullet$ The UV line ratios consistent with metal rich BLR (Sect. \ref{compar}).   The continuum source is hidden  from the view while the BLR is  partially observable.

$\bullet$ The core-ERQ intermediate UV line FWHM  values between  QSO1 and QSO2 at similar $z$ (Sect. \ref{intro}) and the lack of broad wings typical of blue QSO1 (\citealt{Hamann2017}).  This can be explained  if the inner BLR is obscured. The inner BLR clouds would be responsible for the broad wings of the UV lines since they are closer to the SMBH and are expected to have higher velocities (gravitational and non-gravitational due to radiation driven outflows, for instance). In such situation, the FWHM of the broad emission lines is not a reliable tracer of the BLR kinematics and not adequate to estimate black hole masses.

 $\bullet$ The large line-of-sight hydrogen column densities $N_{\rm H}\ga 10^{23}$ cm$^{-2}$ implied by X-ray observations are unlikely due to galactic absorption and are naturally explained by the central AGN dusty obscuring structures (for instance, the classical equatorial optically thick dusty torus and polar dust; \citealt{Ramos2017}, \citealt{Goulding2018}).
Moreover, the $N_{\rm H}$ measured for typical type 1 AGN, for
which a nearly face-on orientation is expected, are overall at least a
factor of ten lower than those measured for core-ERQ (e.g. \citealt{Mateos2005}, \citealt{Mateos2010}, \citealt{Corral2011}).

$\bullet$ The intermediate REW$_{\rm [OIII]}$ between QSO1 and QSO2.
Orientation seems to play a major role  (although it is not the only driver) in the variance of the REW$_{\rm [OIII]}$   in QSO1.  REW$_{\rm [OIII]}$  is used as indicator of the accretion disk inclination (e.g. \citealt{Risaliti2011}, \citealt{Bisogni2017}). Its value increases when we move from face-on to more edge-on systems, due to the  decreasing contribution of  the AGN continuum and  FeII contamination from the BLR.

The [OIII]$\lambda$5007  REW of  \cite{Perrotta2019} sample have REW$_{\rm [OIII]}\sim$42-646  \AA\ with median  177 \AA.   For comparison,    the  sample of WISE/SDSS selected hyper-luminous (WISSH) QSO1 {\it with detected and measurable}   REW$_{\rm [OIII]}\sim$0.2-80 \AA~  (median 2.8 \AA), being the dispersion of these values due at least in part to orientation (e.g. \citealt{Bischetti2017,Vietri2018}).  Another parameter that influences REW$_{\rm [OIII]}$  is $L_{\rm bol}$, with  very luminous systems showing lower REW$_{\rm [OIII]}$ (the Baldwin effect mentioned above).   The $L_{\rm bol}$ of the core-ERQ  is within the range of the WISSH QSO1 ($>10^{47}$ erg s$^{-1}$). If subestimated due to obscuration (\cite{Hamann2017}), the large REW for such high intrinsic $L_{\rm bol}$  further support  a more edge-on inclination than luminous blue QSO1.

$\bullet$ NV/CIV vs. FWHM$_ {\rm CIV}$.

The general properties (ionization, density, kinematics, abundances) of the BLR and the NLR in AGN are extremely different, and this is clearly reflected on the BLR and NLR  line FWHM and the UV emission line ratios (see comparison between the NLRG and the QSO1 line ratios in Fig. \ref{brotherton1}). It is natural to expect that as we move  gradually  from an edge-on to a face-on orientation, the UV emission line properties should  change from QSO2 type to QSO1 type, as the BLR emission becomes more prominent. 

The UV line FWHM and the NV line ratios are two of the most clearly distinct parameters between type 1 and type 2 AGN spectra. As explained in the previous section,  high NV ratios are characteristic of the  BLR, but not of the NLR.  Thus, we naturally expect that FWHM$_ {\rm CIV}$ and NV/CIV both increase from type 2 to type 1 orientation, as the BLR becomes gradually more visible and its contribution to the UV spectrum becomes increasingly dominant relative to the NLR.

\begin{figure}
\centering
\hskip-3mm
\includegraphics[width=0.50\textwidth,height=0.49\textwidth]{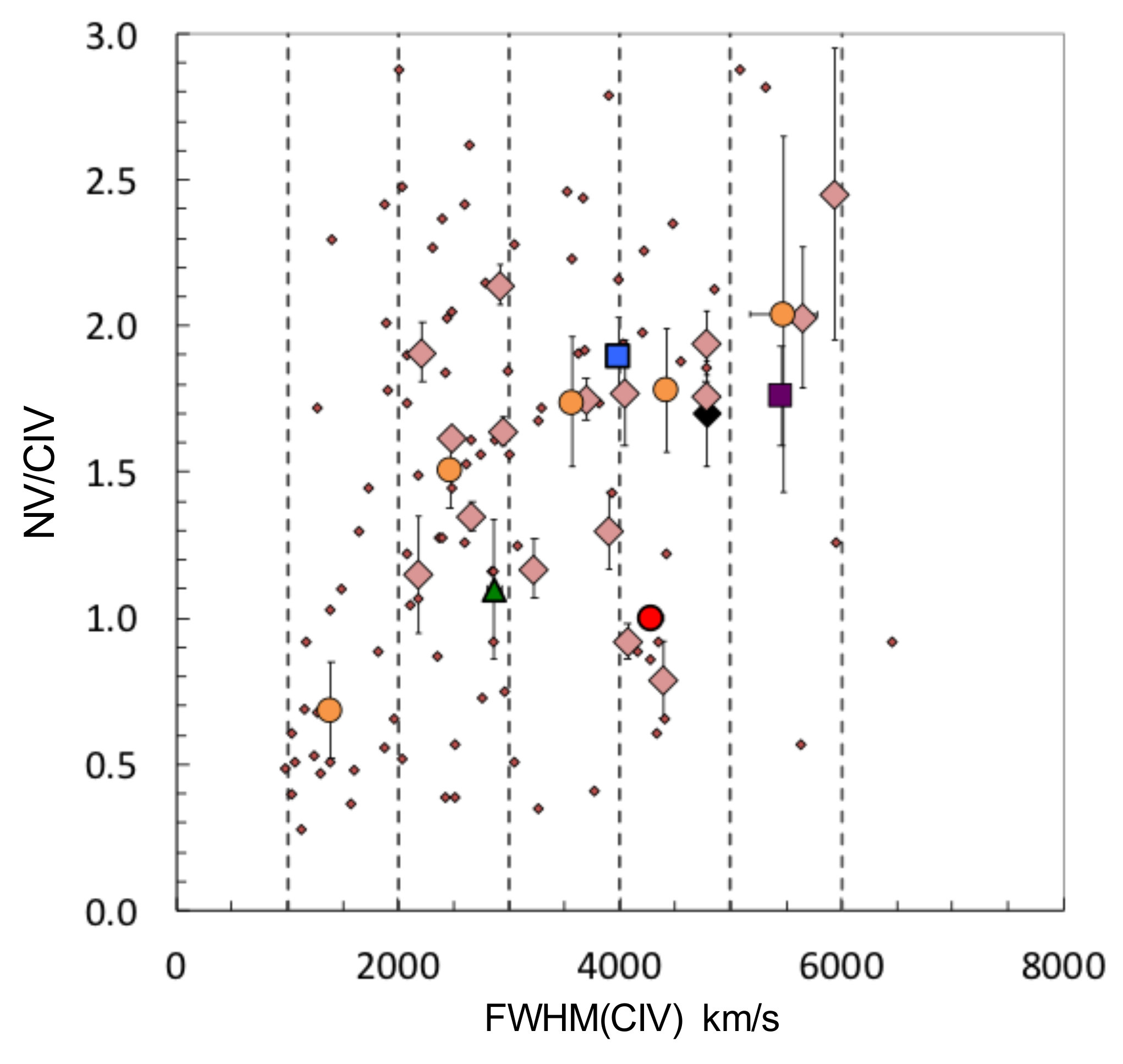}
\caption{NV/CIV vs. FWHM$_{\rm CIV}$ for the 97 core-ERQs in \citet{Hamann2017}.  J1714+4148, the four  core-ERQ from \citet{zak16} and the reamaining core-ERQ in \cite{Perrotta2019} are plotted with the same colour code as in Fig. 1. All other core-ERQ from \citet{Hamann2017} are plotted with small dark red diamonds. The core-ERQ  have been organised in five groups according to the FWHM$_ {\rm CIV}$ range of values, indicated with the vertical dashed lines. The median NV/CIV and FWHM$_ {\rm CIV}$ in km s$^{-1}$ of the five groups are shown with large orange circles. The errorbars correspond to the standard error on the median values for each group. A clear increase is observed of the NV/CIV ratio with line width.}
\label{figham}
\end{figure}

 We have grouped the 97  core-ERQs in \citet{Hamann2017} in five groups according to the range of FWHM$_ {\rm CIV}$: 1000-2000, 2000-3000, 3000-4000, 4000-5000 and 5000-6000 km s$^{-1}$. We have calculated the median value of  FWHM$_ {\rm CIV}$ and NV/CIV for all groups and  plotted them in    Fig. \ref{figham}. The scatter of the whole sample is huge, but  when considering the median values, it is found that as the lines become broader, NV/CIV increases.   Objects with FWHM$<$2000 km s$^{-1}$ maybe similar in orientation to  the QSO2 sample of  \citet{Alexandroff2013},  although $\sim$10 times more luminous.  The large scatter of the data  shows that orientation is not the only parameter influencing NV/CIV and FWHM$_ {\rm CIV}$. Naturally, other parameters such as $n$,  metallicity, ionization parameter $U$ and the nature of the kinematics must be involved (as is the case in other AGN).  In spite of this complexity,  that   the median values behave as expected from a gradual  change of inclination, adds support to the idea that orientation   plays  an important role.  According to this scenario, groups of increasingly broader CIV  have an increasing relative contribution of the BLR to  the emission line fluxes, that would result on  higher NV/CIV,  more typical of QSO1.

\begin{figure}
\centering
\hskip-5mm
\includegraphics[width=0.51\textwidth,height=0.43\textwidth]{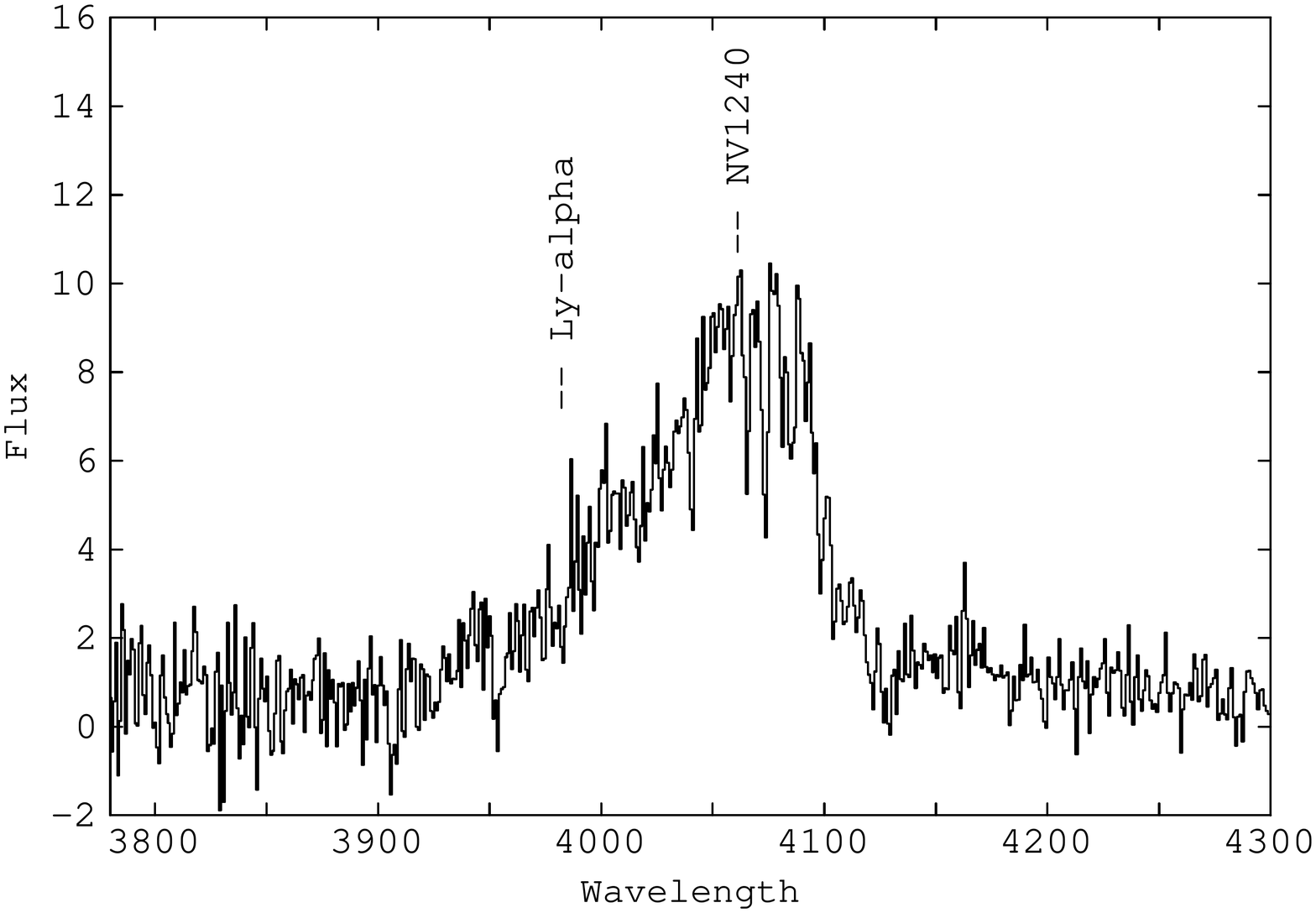}
\vskip-12mm
\hskip-5mm
\includegraphics[width=0.51\textwidth,height=0.43\textwidth]{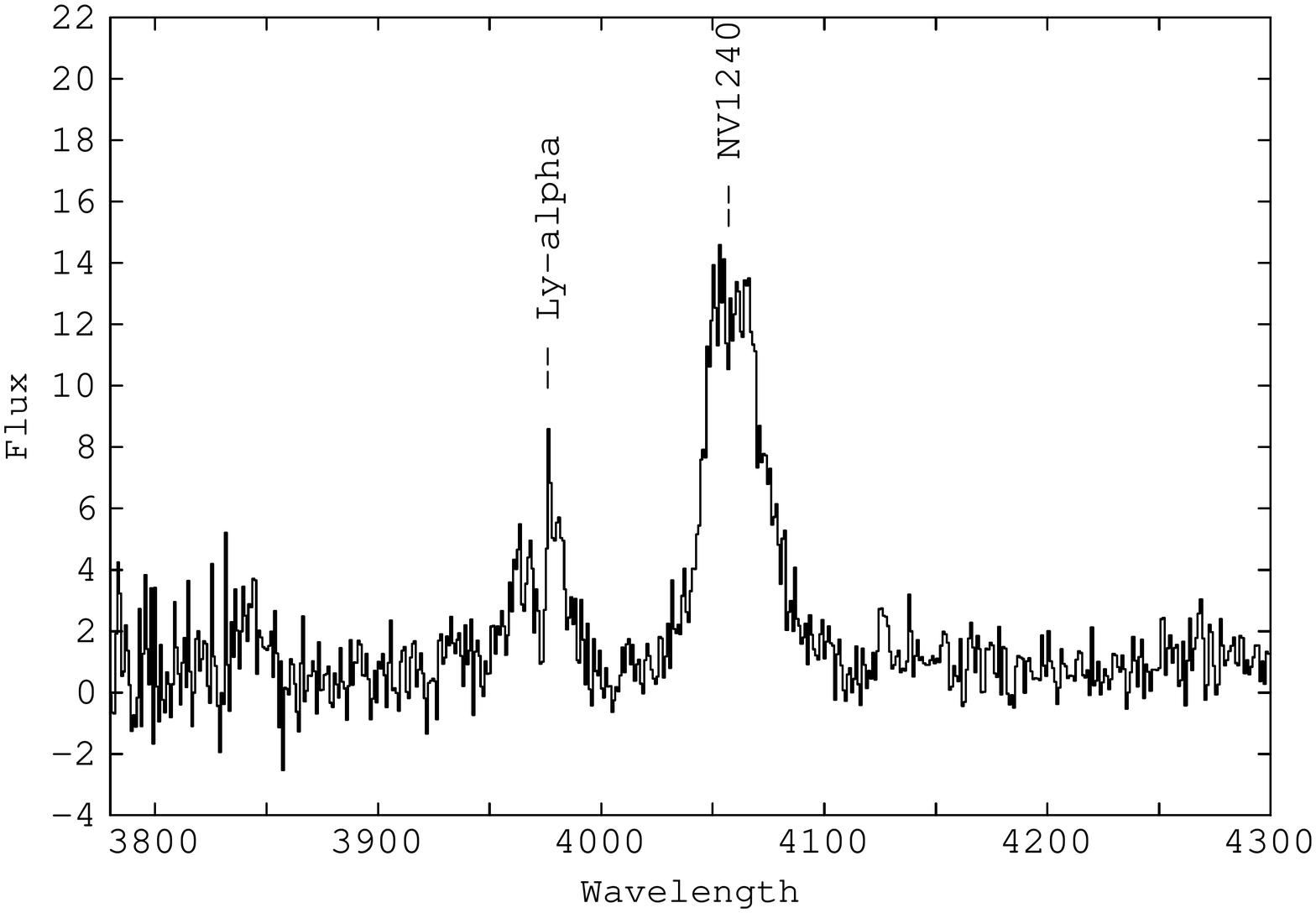}
\vskip-12mm
\hskip-5mm
\includegraphics[width=0.51\textwidth,height=0.43\textwidth]{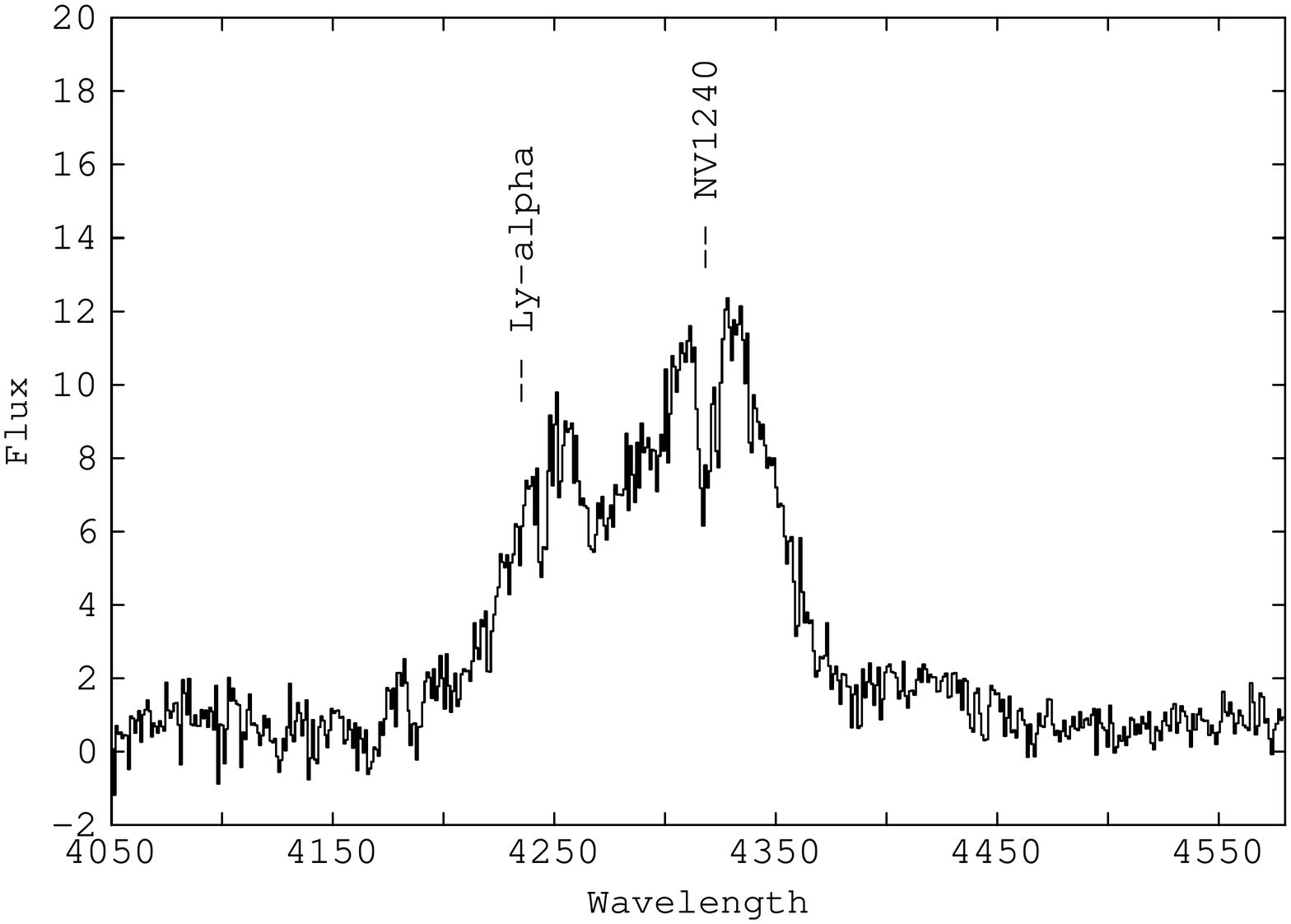}   
\caption{SDSS spectra of  J1031+2903 (top),  J1356+0730 (middle) and J1604+5633 (bottom) in the redshifted  Ly$\alpha$+NV region. Ly$\alpha$ is severely absorbed in many core-ERQ. It shows absorption across thousands of km s$^{-1}$, sometimes both  redshifted and bluesfhited, as in these three objects. The expected Ly$\alpha$ and NV$\lambda$1240 central $\lambda$ are shown based on the CIV and SiIV+OIV] redshifts. Fluxes in units of $\times$10$^{-17}$ erg s$^{-1}$ cm$^{-2}$ \AA$^{-1}$. Wavelength in \AA.}
\label{lya}
\end{figure}

$\bullet$ An intermediate orientation may also explain the heavily absorbed Ly$\alpha$, which suggests the presence of large amounts of absorbing gas. 

 The  Ly$\alpha$ profile in core-ERQ is often absorbed across most or even  the entire spectral line profile (Fig. \ref{lya}; see also \cite{Hamann2017}).  It is likely that in at least  some objects, the emission at $\sim$1216 \AA~ flux is not dominated by Ly$\alpha$ but by OV]$\lambda\lambda$1214,1218 instead  (\citealt{Humphrey2019b}). The Ly$\alpha$ absorbers span a  broad  range of velocities of thousands km s$^{-1}$ which are both blueshifted and redshifted. This is reminiscent of the broad line absorbers (BLA)  (also both blueshifted and/or redshifted depending on the lines)   found by \cite{Zhou2019} in a small sample of QSO1. They propose that the  highly redshifted BAL arising from neutral hydrogen, helium and FeII atoms are due to inflowing gas directly feeding the accretion disk. The authors propose high inclination angles, intermediate between face-on and edge-on and a location for  the absorbers  between the accretion disk and the dusty torus.
 
 Such gas, possibly mixed with dust, may be the same reservoir  behind the illuminated phase of a disk-like  BLR inflated by an accretion disk wind proposed by 
\cite{Czerny2011} and \cite{Baskin2018}.

\subsection{The [OIII] outflows and the effects of orientation}

The unified model proposes  that the [OIII]$\lambda$5007 emission in AGN originates from regions outside the central obscuring structure and should therefore be independent of orientation.  On the other hand, because the AGN continuum is strongly anisotropic, some [OIII] properties do indeed depend on orientation.   QSO1 studies have  shown that not only REW$_{\rm [OIII]}$ changes as a function of orientation, but also  the prominence and observed kinematics of the outflowing gas in the NLR as traced by the [OIII] profile (e.g. \citealt{Shen2014}, \citealt{Bisogni2017}, \citealt{Marziani2017}). The blue broad outflow   component  decreases both in intensity and in velocity  shift relative to the reference wavelength with increasing   REW$_{\rm [OIII]}$. They interpret these results in terms of a gradual change of orientation from face-on (low REW$_{\rm [OIII]}$) to more edge on (high REW$_{\rm [OIII]}$) positions.

In an orientation based scenario, the most extreme kinematics associated with an ionised outflow are expected to be observed in face-on luminous quasars. This means that extreme [OIII] outflows with velocities of several 1000s km s$^{-1}$ should be detected in QSO1  with $L_{\rm bol}$  similar to those of core-ERQs. On the contrary, \cite{Perrotta2019} found that core-ERQ exhibit the broadest and more blueshifted [OIII] emission lines ever reported, with outflow velocities $W_{\rm 90}$ about three times larger than those of luminosity-matched blue quasar samples (see also \cite{zak16}). These extreme and apparently unique [OIII] kinematics contradict the orientation scenario. 

We revise next whether there is enough evidence to sustain this conclusion.

\subsubsection{Fitting [OIII] in luminous QSO1}
\label{oiii}

\cite{Perrotta2019} considered for the comparative QSO1 sample the five WISSH QSOs presented in \cite{Bischetti2017}, and the 74 luminous QSO1 in \cite{Shen2016}, which have 46.2$\le log(L_{\rm [OIII]}) \le$48.2 and 1.5$< z <$3.5.  
They re-built the artificial [OIII]$\lambda$5007 lines for each QSO1, using the best-fit gaussian parameters quoted by \cite{Shen2016} and \cite{Bischetti2017}. They compared the inferred non-parametric velocities $W_{\rm 90}$ with those of core-ERQs. No errors have been estimated. They derived   an average $W_{\rm 90}$= $1550 \pm 300$ km s$^{-1}$ for this QSO1 sample.

The first issue to take into account is that the low  REW$_ {\rm [OIII]}$, the complex H$\beta$+[OIII] blend and the presence of  strong and complex FeII emission in luminous QSO1 makes the determination of [OIII] spectral parameters very challenging.   Different works have shown that, albeit with a very large scatter, REW$_{\rm [OIII]}$ decreases with increasing AGN luminosity (the [OIII] Baldwin effect; \citealt{Brotherton1996}, \citealt{Dietrich2002}, \citealt{Shen2016}).   
This is, at the highest luminosities, the [OIII] lines, if detected, are more difficult (often impossible) to parametrise due to their low REW and the relatively stronger contamination  by the broad H$\beta$ and the underlying FeII multiplets. 
 We argue that the particular orientation of core-ERQ facilitates the detection and parametrisation of  the  outflows, thanks  to the  higher contrast of the [OIII] lines relative to the continuum, the FeII multiplets and to the broad H$\beta$, since these are totally or partially obscured.

Indeed, we note that i) about 70\% of the  18 WISSH QSO1 analysed in \cite{Vietri2018} (see also \citealt{Bischetti2017}) shows weak/absent [OIII] emission, hindering the detection and accurate characterisation of the outflows in most of their targets, and ii) about 35\% of the QSO1 in \cite{Shen2016} is associated with [OIII] detections with $<3\sigma$ significance. Furthermore, the QSO1 in \cite{Shen2016} are characterised by low REW, with an average  REW$_{\rm [OIII]}$ = $14\pm 10\AA$ (see Fig. 4 in \citealt{Shen2016}). Therefore, the outflow velocities must be associated with very large uncertainties.

Recently, \cite{Coatman2019} re-analysed all rest-frame spectra of z>1.5 type 1 QSOs  available from the literature, in order to infer the properties of ionised outflows traced by [OIII] gas with a uniform analysis strategy. The authors provided new independent velocity measurements for the \cite{Shen2016} targets, also reporting significant differences in the derived [OIII] fluxes and velocities. They showed that such discrepancies are not due to the quality of the analysed near-infrared spectra, but to the systematic effects associated with the fit. For instance, the chosen templates to model the FeII emission (see e.g. \citealt{Kovacevic2010}, \citealt{Vietri2018} and refs therein), the adopted function(s) to reproduce the BLR emission (for instnace, single or multiple Gaussian or Lorentzian functions, or broken power-laws; see e.g. \citealt{Nagao2006}, \citealt{Shen2011}), and the shape of the underlying continuum (e.g. \citealt{Mejia2016,Varisco2018}) can significantly affect the analysis results. All these systematic effects are more and more important for QSO1 with high $L_{\rm bol}$, for which we expect more extreme outflows (e.g. \citealt{King2015,Ishibashi2018}) but also higher emission from the more nuclear regions relative to the [OIII] line flux (e.g. \citealt{Shen2014}, \citealt{Shen2016}).

To further highlight the difficulties in deriving robust [OIII] kinematics of high luminosity QSO1, we derive new outflow velocity measurements for a subsample of blue QSO1 presented in \cite{Shen2016}, using a different method, the multicomponent simultaneous fit analysis (e.g. \citealt{Brusa2015}).
We focus on the 13 out of 19 targets at z$>2$ (as the core-ERQs) associated with good quality NIR spectra. We refer to Appendix \ref{AppendixFit} for details regarding the fit analysis and the main differences with respect to the previous work by \cite{Shen2016} and \cite{Coatman2019}. In Table \ref{shen} we report the $W_{80}$ and $W_{90}$ measurements we derived from the [OIII] profiles, together with those from \cite{Shen2016} and \cite{Coatman2019} analysis  (see also Table \ref{OIIIbestfitparams} in Appendix \ref{AppendixFit}).

\begin{figure}
\centering
\vskip0mm
\hskip-2mm
\includegraphics[width=0.50\textwidth,height=0.48\textwidth]{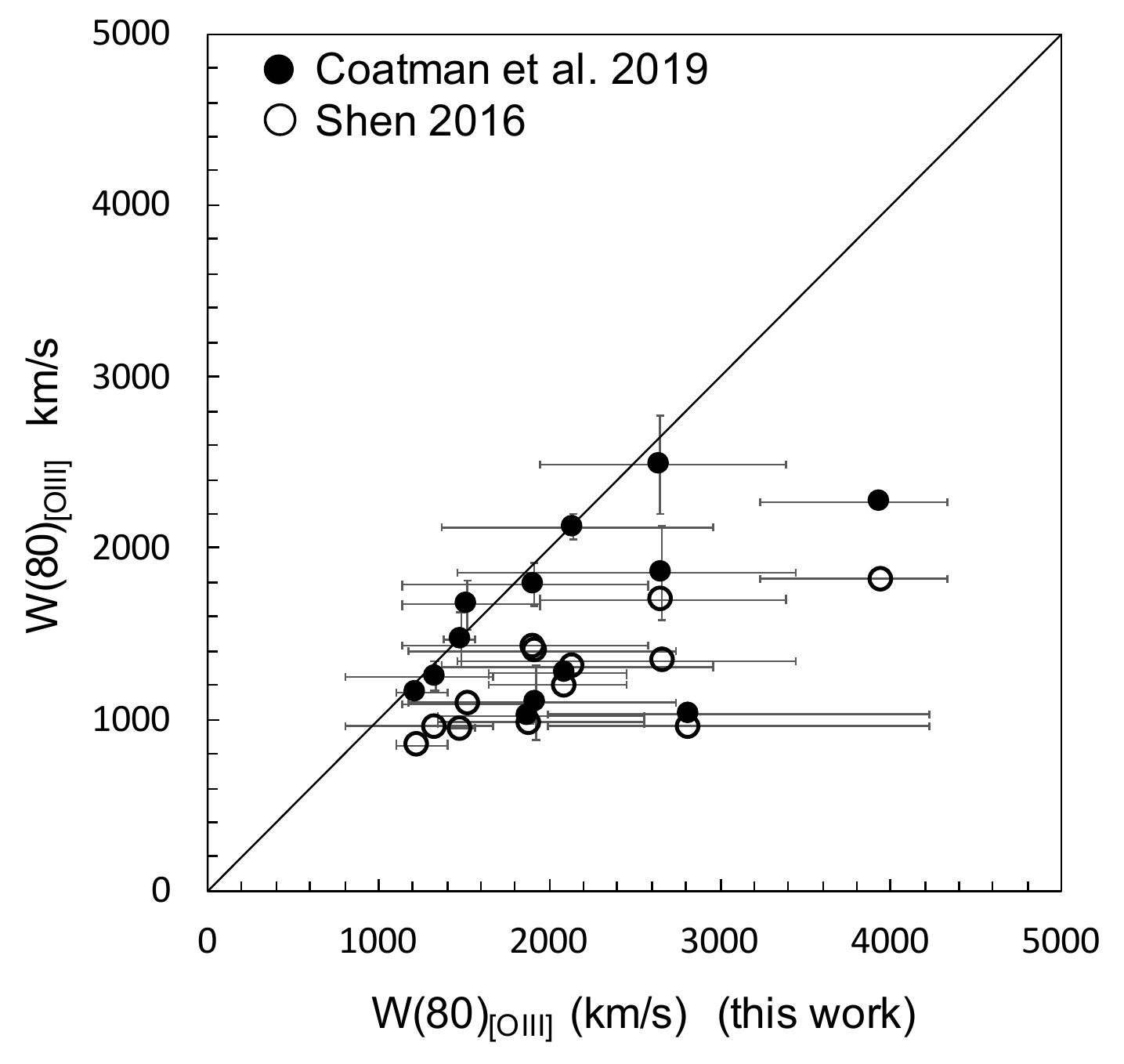}
\caption{Comparison of our [OIII]$\lambda$5007 $W_{\rm 80}$ with  \cite{Coatman2019} and  \cite{Shen2016} for 13 luminous QSO1 at $z>$2. The large error bars and the different results reflect the difficulty to obtain accurate [OIII] kinematic parameters in distant ($z>$2) luminous QSO1.}
\label{compW80}
\end{figure}

The three sets of $W_{\rm 80}$  values are also compared  in Fig. \ref{compW80}. 
Our results are in better agreement with \cite{Coatman2019} than with \cite{Shen2016}, who measures lines 1.3 to 3 times narrower.  The differences in the analysis results between our work and these two studies highlight the difficulties in deriving robust [OIII] kinematic measurements for very luminous QSO1. Such difficulty is reflected also by the large errors of our $W_ {\rm 80}$ values. These are larger than those in \cite{Coatman2019}.  We consider our errors more realistic, since they take into account (whenever possible) all systematic effects related to the modelling of continuum, FeII, BLR and NLR emission (see Appendix \ref{AppendixFit}). Non-parametric velocities errors are not available for \cite{Shen2016}.

The 13 QSO1  discussed above have $L_ {\rm [OIII]}\sim$(0.8-15.0)$\times$10$^{44}$ erg s$^{-1}$. We infer $W_ {\rm 90}\ge$2000 km s$^{-1}$ for 11 of them (see Table \ref{shen}). All are above the median $W_ {\rm 90}=$1568 km s$^{-1}$ quoted by \cite{Perrotta2019} for the entire  QSO1 sample from \cite{Shen2016}. 
Out of the 18 QSO1  within the $L_ {\rm [OIII]}$ range as the subsample fitted by us, they find only three with $W_ {\rm 90}\ge$2000 km s$^{-1}$ (see their Fig. 4). Based on the \cite{Shen2016} fits,  they clearly obtain significantly narrower lines than core-ERQ for the blue QSO1 comparative sample.
With all the caveats mentioned above,  it is clear that  these kinematic parameters  must be affected by very large uncertainties. Unless errors are quoted, any comparison with core-ERQ can only be considered tentative. 

It is also important to note that the $L_{\rm [OIII]}$ core-ERQ values in \cite{Perrotta2019} are lower limits.  They estimated $L_{\rm [OIII]}$ for each object, multiplying the  [O III] REW by the luminosity at 5007 \AA. To obtain the continuum fluxes at this $\lambda$, they linearly interpolated the WISE W1 (3.6$\mu$m) and SDSS $i$ or $z$ magnitudes and then converted  the derived values at  5007  \AA ~ into  fluxes. As pointed out by the authors, the luminosities could be on average 6.3 times higher, since the median core-ERQ spectral energy distribution (SED) is suppressed at $\sim$5000\AA~ by about 2 mag relative to normal/blue quasars (\citealt{Hamann2017}). The correction factor could be even higher (Sect. \ref{regions}).

\begin{figure*}
\centering
\includegraphics[width=1.0\textwidth,height=0.5\textwidth]{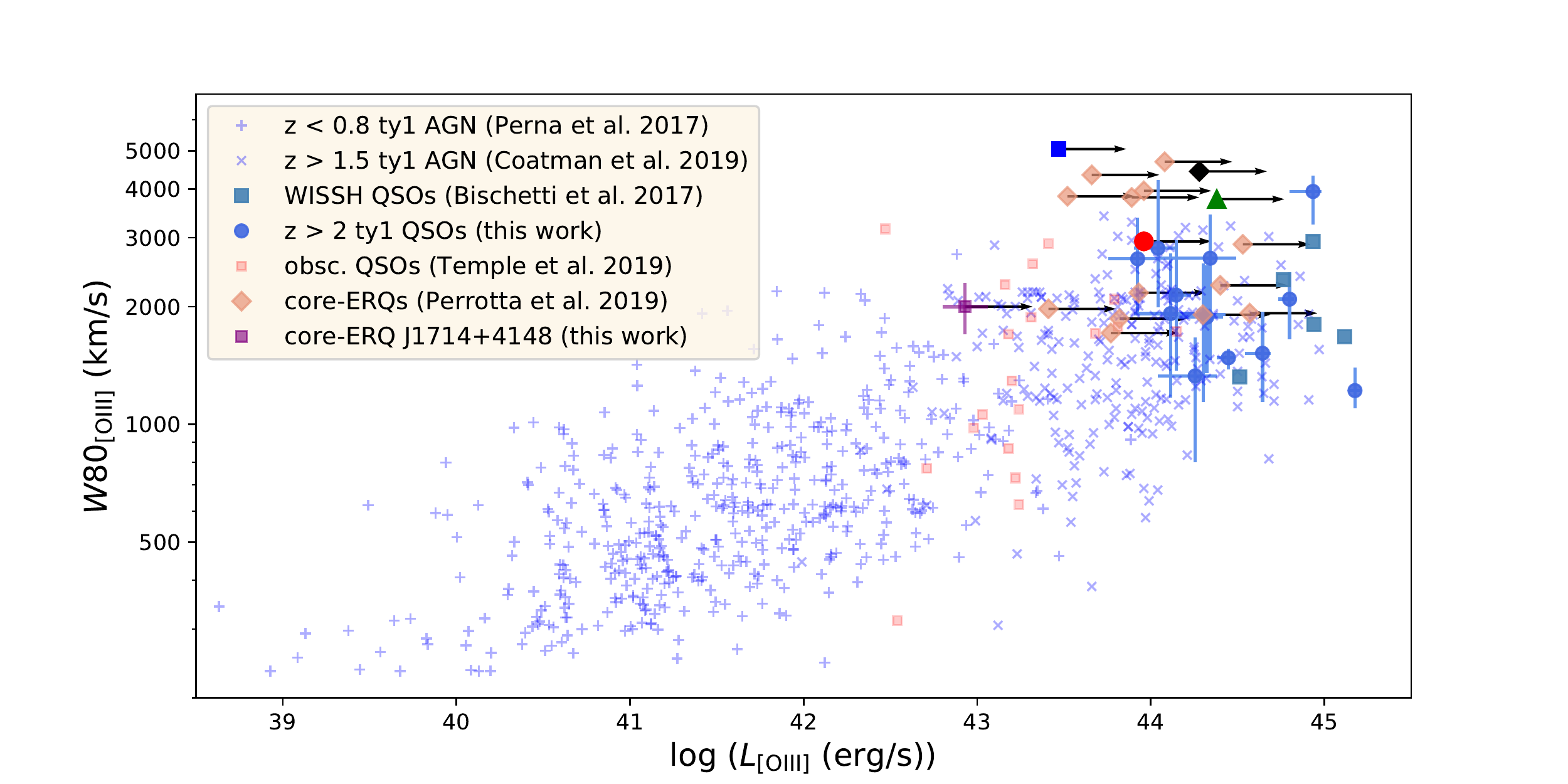}
\caption{Comparison of several type 1 AGN samples in the $W80_{\rm [OIII]}$ vs. $L_{\rm [OIII]}$ plane. There is a clear correlation between $W80_{\rm [OIII]}$  and $L_{\rm [OIII]}$. Core-ERQ fall on it, with or without taking into account an extinction correction. Some QSO1 from other samples show [OIII] as broad as core-ERQ. Morevoer, there is not solid evidence to support that core-ERQ show more extreme kinematics than unobscured QSO1 of similar luminosity.}
\label{LOIII-W80}
\end{figure*}

\subsubsection{[OIII] kinematics. Comparison between core-ERQ and luminous QSO1}
\label{kincomp}

In  Fig. \ref{LOIII-W80} we compare the [OIII] kinematics of core-ERQs as traced by $W_ {\rm 80}$ with other samples of type 1 AGN. In particular, we consider:

$\bullet$  \cite{Perrotta2019} core-ERQ (the four objects from \citealt{zak16} are highlighted with different symbols as in Fig. 2). $W_{\rm 80}$ have been derived rescaling  their $W_{90}$, assuming $W_{80}\sim$0.72$\times W_{80}$.  This  is consistent with the value expected for a gaussian profile, and with the $W_{\rm 80}/W_{\rm 90}$  ratios in \cite{Zakamska2014} for type 2 AGN, in \cite{zak16} for core-ERQs and in \cite{Coatman2019} for QSO1;

$\bullet$ SDSS J1714+4148 (see Sect. \ref{observations});

$\bullet$ 13 QSO1 from \cite{Shen2016}  at $z>$2 using the result from our new fits;

$\bullet$ Low $z$ X-ray detected SDSS type 1 AGN at $z<$0.8 from \cite{Perna2017a}.  This sample allows us to highlight the correlation between $L_{\rm [OIII]}$ and $W_{80}$ over a range of five dex, as well as its large scatter at fixed luminosity;

$\bullet$ Five WISSH QSO1  from \cite{Bischetti2017}.  We have measured $W_{\rm 80}$ by reconstructing the [OIII] profiles with the Gaussian parameters provided by the authors in their tables;

$\bullet$ \cite{Coatman2019}  sample of luminous QSO1   (45.5$<log(L_{\rm bol})<$49.0) at  1.5$<z<$4.0. The 13 QSO1 at $z>$2  re-analysed in this work have not been excluded from the figure (see Fig.\ref{compW80}  and Table \ref{shen}  for a comparison between their fit results and ours);

$\bullet$ 22 heavily reddened quasar candidates  from the UKIDSS-LAS, VISTA VHS and VIKING imaging surveys at 2.0$\la z\la$2.6 and 46.0$\la$log($L_{\rm bol}$)$\la$48.5 (\citealt{Temple2019}).

These are the main conclusions:

- core-ERQ fall on the $W_ {\rm 80}$ vs. $L_ {\rm [OIII]}$ correlation defined by different samples of type 1 AGN. This is consistent with the scenario in which more kinematically extreme outflows are triggered by  more luminous AGN. Any reasonable correction factor to $L_ {\rm [OIII]}$ due to dust extinction  would maintain them within the correlation.

- there are very luminous non-obscured QSO1 with [OIII] as broad as  core-ERQ (\citealt{Shen2016}, \citealt{Bischetti2017}, \citealt{Coatman2019}).   9 out of the 22 heavily reddened quasar candidates (\citealt{Temple2019}) show  [OIII] kinematics similar to  some core-ERQ (see Fig. \ref{LOIII-W80}).  The extreme  [OIII] kinematics is, therefore, not exclusive of this object class.

- taking into account the lower limits on $L_ {\rm [OIII]}$ and the new  results on the blue QSO1 sample of \cite{Shen2016} it cannot be claimed that   core-ERQ show in general more extreme kinematics compared with blue QSO1 of matched bolometric luminosities.

- Extreme [OIII] kinematics is indeed much more frequent in core-ERQ ($\sim$100\%) than in QSO1, as already pointed out by \cite{Perrotta2019}.  Considering all AGN samples, out of the  9 objects with the most extreme kinematics ($W_{80}\ga$4000 km s$^{-1}$), 8 are core-ERQ. Whether this is a real intrinsic difference due to truly more extreme outflows being triggered in core-ERQ or rather an artificial effect due to uncertainties resultant of  all caveats mentioned above     and the easier detection and parametrisation of outflows in core-ERQ is unknown.

Based on the above, we conclude that the extreme [OIII] kinematics of core-ERQ do not pose a problem for the orientation based scenario.
\begin{table*}
\centering
\tiny
\begin{tabular}{llllllllllllllll}
\hline
Name	&	$z_{\rm sys}$	&	$L_{\rm[OIII]}$ $\times$10$^{44}$ &	$W_{\rm 80}$  & $W_{\rm 90}$  & $L_{\rm[OIII]}^{C19}$ $\times$10$^{44}$ &	$W_{\rm 80}^{C19}$  &  $W_{\rm 90}^{C19}$	 & 	$W_{\rm 80}^{Sh16}$  &  $W_{\rm 90}^{Sh16}$ 	\\ 
	&		&	erg s $^{-1}$	 & km s$^{-1}$ & km s$^{-1}$  & erg s$^{-1}$  &   km s$^{-1}$ & km s$^{-1}$  &  km s$^{-1}$ & km s$^{-1}$   \\ \hline
J0149+1501 	&	2.0726  	& 		1.3$_{-0.5}^{+0.5}$	& 1920$_{-750}^{+820}$  & 		2540$_{-1020}^{+1090}$   & 1.0	& 1100$\pm$221 & 	1457$\pm$287 &	1400 & 1695	 \\
J1421+2241 	&	2.1887	 	& 	2.0$_{-0.7}^{+0.7}$  &	1910$_{-770}^{+670}$	& 	 	2500$_{-990}^{+890}$ 	& 2.0 &	1784$\pm$126	& 2341$\pm$178	& 1430 &	1750 \\
J1431+0535	& 	2.1004		& 		2.2$_{-1.1}^{+0.9}$ &	2660$_{-1200}^{+780}$	 & 3560$_{-1580}^{+1100}$	&  1.9 &  1856$\pm$275 &	2410$\pm$320  &	1340	& 1720	 \\
J1436+6336  	& 	2.0665		& 		2.1$_{-0.7}^{+0.5}$  &  	1880$_{-530}^{+670}$	&	2480$_{-740}^{+880}$ & 1.1  & 	1019$\pm$45	  & 1304$\pm$63 & 	980	 &1270	\\
J1220+0004 	& 	2.0479		& 	1.1$_{-0.3}^{+0.3}$  &	2820$_{-830}^{+1400}$	 & 3690$_{-1160}^{+1890}$	& 0.3 & 	1033$\pm$39	& 1323$\pm$88	& 960	& 1260   \\
J0250-0757 	& 		3.3376		& 	1.8$_{-0.7}^{+0.6}$ &  	1330$_{-530}^{+340}$	& 1800$_{-780}^{+570}$  & 1.8 & 	1250$\pm$85	& 1740$\pm$106	& 960	& 1360	\\
J0844+0503 	& 		3.3603		& 	8.6$_{-2.3}^{+1.0}$ & 3940$_{-700}^{+390}$	&  	4960$_{-890}^{+560}$	 & 49.0 & 2270$\pm$7	& 2912$\pm$8 &	1820 & 	2250 \\
J0942+0422 		& 	3.2790	& 	6.3$_{-0.9}^{+0.4}$ & 2090$_{-440}^{+360}$ &	3140$_{-950}^{+860}$    &  5.1 & 	1266$\pm$23	 & 1616$\pm$35	& 1200	& 1520 \\	J0304-0008 		& 	3.2859		& 	15.0$_{-1.0}^{+1.0}$  	&  	1220$_{-120}^{+179}$ 	&	1920$_{-270}^{+230}$   & 8.1 &  1155$\pm$64	& 1634$\pm$129 & 	850 & 	1140 \\	
J0843+0750 		& 	3.2648		& 	2.8$_{-0.4}^{+0.2}$ &	1480$_{-100}^{+80}$ &	2060$_{-145}^{+110}$ & 	2.6 & 1464$\pm$163 &	2008$\pm$273	 &950 &	1210  \\
J1019+0254 		& 	3.3829		& 	1.4$_{-0.1}^{+0.3}$  & 	2140$_{-770}^{+820}$ &		2870$_{-1070}^{+970}$ &  0.7 &  	2121$\pm$74 &  2723$\pm$96	 & 1310	& 1670 \\
J0259+0011 		& 	3.3724		& 	4.4$_{-0.9}^{+0.5}$ & 1520$_{-380}^{+420}$ 	&	2070$_{-570}^{+650}$  & 3.4 & 1668$\pm$141	& 2306$\pm$223	& 1090	& 1490 \\
J1034+0358 		& 	3.3918		& 	0.84$_{-0.27}^{+0.27}$ & 	2650$_{-710}^{+730}$ 	&	3430$_{-920}^{+930}$ & 0.5 &  2484$\pm$287	& 2959$\pm$305 & 	1700 &	2060	\\
\hline
\end{tabular}
\caption{QSO1 in \cite{Shen2016} at $z\ge$2 for which [OIII] could be fitted. The [OIII]  $W_{\rm 80}$ and $W_{\rm 90}$ values  inferred by us, \cite{Coatman2019} and \cite{Shen2016} are compared. $z_{\rm sys}$ was measured relative to the peak of [OIII]$\lambda$5007. According to our fits,  most of these QSO1 show  $W_{90}$ in the range measured for core-ERQ.}
\label{shen}
\end{table*}

\subsubsection{Kinetic power of the [OIII] outflows}

High outflow kinetic powers of at least $\sim$3-5\% of $L_{\rm bol}$ have been inferred for core-ERQ (\citealt{Perrotta2019}). In principle, such powerful winds   have the potential to affect the evolution of the host galaxies.  For these calculations, sizes $R$=1 kpc and densities $n$=200 cm$^{-3}$ were assumed.  The authors provide a range of uncertainties to account for a possible range $n\sim$100-1000 cm$^{-3}$ and outflow sizes $R\sim$0.5-7 kpc. 

The most likely situation is that the ionized outflows are spatially unresolved (\citealt{Karouzos2016}, \citealt{Villar2016}, \citealt{Husemann2016}, \citealt{Rose2018}, \citealt{Tadhunter2018}). The integrated spectra used for the outflow measuments are expected to be dominated  by the compact  NLR where a broad range densities of up to  $n\sim$10$^7$ cm$^{-3}$ are likely to exist (see \citealt{Villar2016} for a discussion). For a broad distribution of densities, the
emission of a certain forbidden line is expected to peak at gas with
$n$ similar to its critical density $n_{\rm crit}$ (e.g. \citealt{ste14}). For the [OIII]$\lambda\lambda$4959,5007 lines, $n\sim n_{\rm crit}$=8.6$\times$10$^5$ cm$^{-3}$. If as argued in Sect. \ref{regions} radiation pressure confinement defines the gas density spatial profile,  we can infer the size of the [OIII] region for eacth object using equation [eq. 1], $n_ {\rm crit}$ and $L_ {\rm ion}$. The inferred sizes are $\sim$73-227 pc, depending on the object. Using these sizes and $n\sim n_ {\rm crit}$, the kinetic powers of the outflows are reduced by a factor of $\sim$300-977 depending on the object and become  $\ll$1\% $L_{\rm bol}$.  The existence of lower $n$ gas in the outflows is of course possible (e.g. \citealt{Baron2019,Perna2017b}) and   it may carry a significant amount of mass, but with the existing data, nothing can be said about it. 

High spatial resolution spectroscopy  is essential to resolve the ionized outflows and characterise more accurately their  sizes, geometry, and density distribution. In the meantime, the     kinetic powers are too uncertain to infer any useful conclusion regarding their potential to affect the evolution of the host galaxies.

\section{Summary and conclusions}
\label{conclusions}

Core-extremely red quasars (core-ERQ) have been proposed to represent an intermediate evolutionary phase in which a heavily obscured quasar is blowing out the circumnuclear ISM with very energetic outflows prior to becoming an optical quasar, as well as sites of extreme  large scale  ($\ga$1 kpc) outflows that inhibit star formation.
 
Based on the  revision of the general UV and optical  emission line properties of  core-ERQ  at $z\sim$2-3 we propose that at least a high fraction of core-ERQ  are very luminous but otherwise normal quasars viewed at an intermediate orientation between type 1 (face-on) and type 2 (edge-on). The general properties are not necessarily indicative of the particular evolutionary  phase mentioned above, but are instead strongly determined by orientation and the effects of the high bolometric luminosities.  

The calculations and results presented here rest on the assumption that core-ERQ  follow many correlations identified in other AGN samples.  Although this is uncertain, the argumentations suggest that they host large broad line regions ($r_{\rm BLR}\sim$several pc). The highly inclined line of sight is such that we observe only part (the outskirts) of the super-solar metallicity BLR. We propose that the inner BLR and the continuum source remain hidden by the torus. 

An intermediate orientation is consistent with many properties of core-ERQ: the UV line ratios similar to  N-loud QSO1, the large  CIV$\lambda$1550 and [OIII]$\lambda$5007  rest-frame equivalent widths compared with QSO1 of similar $L_ {\rm bol}$, the intermediate FWHM of the UV  lines between  QSO1 and QSO2 at similar $z$  and their wingless profiles. It is also consistent with the large $N_{\rm H}\ga$10$^{23}$ cm$^{-2}$  implied by X-ray observations and the heavily absorbed Ly$\alpha$ profiles.

 We present a revised comparison of the [OIII] kinematics in 21 core-ERQ (20 from \cite{Perrotta2019} and SDSS J1714+4148, based on GTC EMIR near infrared  spectroscopy) with other samples of type 1 AGN. Core-ERQ fall on the $W_ {\rm 80}$ vs. $L_ {\rm [OIII]}$ correlation defined by different samples of type 1 AGN. This is consistent with the scenario in which more kinematically extreme outflows are triggered by  more luminous AGN. 

The extreme [OIII] kinematics of core-ERQ are not exclusive of this object class, as demonstrated by  our revised spectral  fits of luminous non-obscured QSO1 at similar $z$ and other recent works (\citealt{Bischetti2017}, \citealt{Coatman2019}, \citealt{Temple2019}).    Contrary to other works (\citealt{Perrotta2019}), we find that there is no evidence to  support that   core-ERQ show in general more extreme kinematics compared with blue QSO1 of matched bolometric luminosities. The difficulty to detect and/or the large uncertainties affecting the parametrisation  of the  [OIII] profiles  in  the luminous QSO1 comparison sample, and the fact that both $L_{\rm bol}$ and $L_{\rm [OIII]}$ of core-ERQ are very uncertain (and  underestimated), lie at the core of this discrepancy.

Extreme [OIII] kinematics are indeed much more frequent in core-ERQ ($\sim$100\%), as already pointed out by other authors (\citealt{Perrotta2019}).    Whether this is a real intrinsic difference due to truly more extreme outflows being triggered in core-ERQ or rather an artificial selection effect due to the difficulties to detect and characterise [OIII] outflows in luminous QSO1 in comparison with the  much  easier detection and parametrisation  in core-ERQ is unknown.

 High spatial resolution spectroscopy  is essential to resolve the [OIII] outflows and characterise more accurately their sizes, geometry and density distribution. In the meantime, the     kinetic powers are too uncertain to infer any useful conclusion regarding their potential to affect the evolution of the host galaxies. 

 As proposed by other authors, core-ERQ are ideal systems to detect and  characterise the most extreme AGN outflows. Whether this is a consequence of the blow-out quasar phase or of a particular intermediate orientation remains to be disentangled.

\begin{acknowledgements}

We thank Jos\'e Acosta Pulido for valuable scientific discussions and the anonymous referee for  the thorough revision of the manuscript. Thanks also to Yue Shen for making his QSO1 NIR spectra available.

Partly based on observations made with the GTC  telescope and the EMIR instrument at the Spanish Observatorio del Roque de los Muchachos of the Instituto de Astrof\'\i sica de Canarias (program GTC14-19A).  We  thank the GTC staff for their support with the observations. We thank Joel Vernet for providing the high $z$ NLRG composite spectrum.

 MVM acknowledges  support  from the Spanish Ministerio de  Ciencia, Innovaci\'on y Universidades  through the grants  AYA2015-64346-C2-2-P and  PGC2018-094671-B-I00 (MCIU/AEI/FEDER,UE). 
MP is supported by the Programa Atracci\'on de Talento de la Comunidad de Madrid via grant 2018-T2/TIC-11715. 
MP acknowledges support from the Spanish Ministerio de Econom\'ia y Competitividad through the grant ESP2017-83197-P.
AH acknowledges FCT Fellowship SFRH/BPD/107919/2015; Support from European Community Programme (FP7/2007-2013) under grant agreement No. PIRSES-GA-2013-612701 (SELGIFS); Support from FCT through national funds (PTDC/FIS-AST/3214/2012 and UID/FIS/04434/2013), and by FEDER through COMPETE (FCOMP-01-0124-FEDER-029170) and COMPETE2020 (POCI-01-0145-FEDER-007672). S.M. acknowledges financial support through grant AYA2016-76730-P  (MINECO/FEDER).
 This research has made use of: 1) the VizieR catalogue access tool, CDS,
 Strasbourg, France. The original description of the VizieR service was
 published in Ochsenbein et al. A\&AS, 143, 23;   2) data from Sloan Digital Sky Survey. Funding for the SDSS and SDSS-II has been provided by the Alfred P. Sloan Foundation, the Participating Institutions, the National Science Foundation, the U.S. Department of Energy, the National Aeronautics and Space Administration, the Japanese Monbukagakusho, the Max Planck Society, and the Higher Education Funding Council for England. The SDSS Web Site is http://www.sdss.org/; 3) the Cosmology calculator by \cite{Wright2006};
4) the NASA/IPAC Extragalactic Database (NED), which is operated by the Jet Propulsion Laboratory, California Institute of Technology, under contract with the National Aeronautics and Space Administration.

\end{acknowledgements}

\begin{appendix}
\label{appendix}

\section{UV line ratios of core-ERQ}
\label{appendix1}

We show in Tables \ref{ratios2} and \ref{ratios3} the line ratios for the 16 core-ERQ not included in Table \ref{ratios}. For all objects, the line fluxes were measured by integrating the flux within the area defined by the line profiles. This method is more  accurate than fitting Gaussian profiles when the lines are highly asymmetric due to, for instance, absorption. Gaussian fits were also applied  when the line of interest was blended with a neighbouring line. This was useful, for instance, to measure  CIII]$\lambda$1909 more precisely. This line is often blended with SiIII]$\lambda\lambda$1883,1892. In a minority of objects, when the lines are very broad,  AlIII]$\lambda$1857 can also contaminate this blend.
In such cases, different constraints were applied to get a range of possible CIII] fluxes. As an example, when necessary, the FWHM of CIII] was fixed in km s$^{-1}$ to be the same as CIV and/or SiIV, taking into account the additional broadening because it is a doublet (1907 and 1909 \AA).  The error bars take into account the range of fits that are acceptable for the different constraints applied.

\begin{table*}
\centering
\small
\begin{tabular}{lllllllllllll}
\hline
Object &	 CIV/CIII] &  CIV/HeII & CIII]/HeII & NV/CIV & NV/HeII &  NIII]/CIII] &  NIV]/CIV   \\	\hline
 J0006+1215 &   $\ga$1.8   	& 3.4$\pm$0.5	 & $\la$1.9	& 2.0$\pm$0.2	     & 7.0$\pm$1.0  &   N/A  & $\la$0.46	   \\ 
J0209+3122 & 1.5$\pm$0.3  & $\ga$3.3 & $\ga$2.3  &   1.2$\pm$0.1 &  $\ga$3.9 & $\la$0.77 &   $\la$0.26 \\  
J0805+4541  &  1.7$\pm$0.3	& $\ga$2.8	 &	$\ga$1.6  &	 	1.8$\pm$0.2 & $\ga$4.9  & $\la$0.68 & $\la$0.34 \\ 
 J0826+0542  & 4.0$\pm$0.3  & 8.2$\pm$0.5 & 2.0$\pm$0.02 &  1.62$\pm$0.02 & 13.2$\pm$0.9 &  $\la$0.41 &  $\la$0.07 \\ 
J0832+1615  &  2.6$\pm$0.3 & 9.8$\pm$2.0  &  3.8$\pm$0.8 &  1.3$\pm$0.1 & 7.4$\pm$1.0  & $\la$0.65 & $\la$0.14  \\ 
J0913+2344 &  4.5$\pm$0.6 & 5.1$\pm$0.7 & 1.1$\pm$0.2 & 1.35$\pm$0.05  &  6.9$\pm$0.9 & $\la$0.86 &  $\la$0.15 \\ 
J0932+4614 &  2.9$\pm$0.3 & 5.6$\pm$0.7 & 2.2$\pm$0.3 & 1.2$\pm$0.2  & $\ga$6.1  &  $\la$0.54 &  $\la$0.16 \\ 
J0958+5000 & 2.3$\pm$0.2  & 11.3$\pm$1.1 & 4.9$\pm$0.6 &  0.8$\pm$0.1 & 8.8$\pm$1.7 & $\la$0.17 &  $\la$0.06 \\ 
J1013+3427 & 2.3$\pm$0.3  & 5.0$\pm$0.8 & 2.2$\pm$0.5  &  0.92$\pm$0.06 & 4.6$\pm$0.8 &  $\la$0.15 &   $\la$0.07 \\ 
J1025+2454 &   2.8$\pm$0.8  &  $\ga$4.9   & $\ga$1.7 &  2.5$\pm$0.5 & $\ga$9.5 & $\la$0.58 & $\la$0.19  \\ 
J1031+2903 &   3.5$\pm$0.3 & $\ga$5.3	& $\ga$1.5 &  1.76$\pm$0.05 & $\ga$9.3 &  $\la$0.73 &  $\la$0.17 \\ 
J1138+4732 & 3.9$\pm$0.5  & 5.2$\pm$0.6 & 1.4$\pm$0.2 &   1.75$\pm$0.07  & 9.2$\pm$1.0 & $\la$0.86  & $\la$0.22  \\ 
J1217+0234 &  6.1$\pm$1.2 & $\ga$6.4 & $\ga$1.0 & 1.64$\pm$0.05  &  $\ga$10.4 & $\la$1.1 & $\la$0.14 \\ 
J1356+0730 & 3.1$\pm$0.4  &  $\ga$5.2  & $\ga$1.7 & 1.91$\pm$0.08  & $\ga$10.0 & $\la$0.65 & $\la$0.19  \\ 
J1604+5633 & 5.1$\pm$0.4  & $\ga$5.0 & $\ga$1.0 &  1.9$\pm$0.1 &  $\ga$9.7 &  $\la$1.4 & $\la$0.20   \\ 
J1652+1728 & 3.6$\pm$0.3  & 6.8$\pm$0.6 & 1.9$\pm$0.2 & 2.14$\pm$0.07  &  14.5$\pm$1.3& $\la$0.18 & $\la$0.06   \\ 
\hline 
\end{tabular}
  \caption{UV line ratios of the 16 core-ERQ  not included in Table \ref{ratios}. This and next table include different ratios. }
\label{ratios2}
\end{table*}

\begin{table*}
\centering
\small
\begin{tabular}{llllllllllllllllll}
\hline
Object 	    & Ly$\alpha$/HeII & Ly$\alpha$/CIV  &   (SiIV+OIV])/CIV  & CII/CIV &  OIII]$\lambda$1663/CIV    \\	\hline
 J0006+1215&     7.4$\pm$1.4	 & 2.2$\pm$0.3 &   0.64$\pm$0.05  & $\la$0.75 &   $\la$0.29    \\	
J0209+3122 &   $\ga$11.2 & 3.4$\pm$0.2 & 0.49$\pm$0.06 & $\la$0.36   &   $\ga$0.30  \\ 
 J0805+4541  &  $\ga$4.6 &   1.7$\pm$0.2  &  0.53$\pm$0.07 &  $\la$0.23 & $\la$0.36    \\ 
 J0826+0542  &   17.5$\pm$0.2 & 2.14$\pm$0.03  &  0.46$\pm$0.01   & $\la$0.10 &  $\la$0.12  \\ 
J0832+1615  & 7.6$\pm$0.2  &  1.4$\pm$0.1 & 0.35$\pm$0.04 &  $\la$0.19 &   $\la$0.18  \\ 
J0913+2344 & 5.8$\pm$0.2  &  1.12$\pm$0.04 & 0.38$\pm$0.03 & $\la$0.19  & $\la$0.15     \\ 
J0932+4614 & $\ga$33   & 6.2$\pm$0.5 & 0.30$\pm$0.03 &  $\la$0.14  & $\la$0.19   \\ 
J0958+5000 & 31.3$\pm$0.3  & 2.8$\pm$0.1 &  0.19$\pm$0.01 &    $\la$0.04 &  $\la$0.09     \\ 
J1013+3427 & 21.6$\pm$0.7  & 4.3$\pm$0.3 & 0.28$\pm$0.02 &  $\la$0.08 &  $\la$0.21 \\ 
J1025+2454 & $\ga$7.2 & 1.5$\pm$0.5 &   0.9$\pm$0.07 &  $\la$0.24 &  $\la$0.21  \\ 
J1031+2903 &  $\ga$2.0 & 0.37$\pm$0.05 & 0.68$\pm$0.03 &    $\la$0.39 & $\la$0.20 \\ 
J1138+4732 &  6.3$\pm$0.1 & 1.20$\pm$0.07 & 0.45$\pm$0.04 & $\la$0.27  &   $\la$0.20 \\ 
J1217+0234 & $\ga$9.9  &  1.55$\pm$0.04 & 0.38$\pm$0.02  &  $\la$0.10  &  $\la$0.16  \\ 
J1356+0730 & $\ga$3.1  & 0.59$\pm$0.05 & 0.55$\pm$0.03 &  $\la$0.24  &   $\la$0.19 \\ 
J1604+5633 &  $\ga$4.7 &   0.95$\pm$0.09 &  0.61$\pm$0.03 &  $\la$0.12 & $\la$0.21   \\ 
J1652+1728 & 6.78$\pm$0.09  &  1.0$\pm$0.04 & 0.37$\pm$0.02 & $\la$0.05  &  $\la$0.15    \\ 
\hline 
\end{tabular}
  \caption{More UV line ratios of the 16 core-ERQ  not included in Table \ref{ratios}.}
\label{ratios3}
\end{table*}

\section{Multicomponent simultaneous spectral fit}
\label{AppendixFit}

\vspace{0.2cm}

 Obtaining accurate [OIII] kinematic parameters is challenging in luminous QSO1 (see Sect. \ref{oiii}). To evaluate  the typical uncertainties on the derived [OIII] velocities, we have reanalysed the 14 out of 19 QSO1 in \cite{Shen2016}  sample at z$>2$ (as the core-ERQs) associated with good quality NIR spectra.  The remaining objects  could not be fitted due to the weakness or absence of the [OIII] lines  in the spectra. 

We used the multicomponent simultaneous fit technique, generally adopted in the literature to reduce the degeneracies between FeII, BLR and NLR emission (e.g. \citealt{Brusa2015,Perna2017a,Perna2017b}). We fitted simultaneously the H$\beta$+[OIII]+FeII complex together with the H$\alpha$+[NII]+[SII] system for the sources at z $\sim 2$, and MgII+FeII+FeIII and H$\beta$+[OIII]+FeII regions for the targets at higher $z$. The best-fit results are shown in Figs. \ref{fitsshen1}, \ref{fitsshen2}, \ref{fitsshen3} and in Tables \ref{shen}, and \ref{OIIIbestfitparams}.
These figures also show the comparison of the [OIII] best-fit obtained by \cite{Shen2016} with ours (we could not do the same for \citealt{Coatman2019} best fit, because [OIII] Gaussian parameters are not tabulated in their paper).  The  spectrum of  J2238-0821 is also shown in Fig. \ref{fitsshen2}. Unlike in \cite{Shen2016}, our fits indicate that NLR emission  is undetected in this source (consistent with \citealt{Coatman2019}).

\subsection{Modelling rest-frame optical spectra}
The H$\alpha$+[NII]+[SII] and H$\beta$+[OIII]+FeII systems of J0149+1501, J1421+2241, J1431+0535, J1436+6336 and J1220+0004 are redshifted in the K- and H-band, respectively. We fitted simultaneously the continuum and all emission lines to reduce the degeneracies between BLR, NLR and FeII emission. In short, we fitted a (single) power-law continuum, an optical FeII template 
(\citealt{Kovacevic2010}) and a combination of Gaussian functions to model BLR and NLR emission lines.

In particular, depending on the complexity of FeII emission, we considered one or two FeII templates (see e.g. \citealt{Vietri2018}). These are convolved with a Gaussian; the width of this Gaussian, the amplitude normalisations and velocity offset of the FeII templates are free variables in our fit (see \citealt{Perna2017a} for further details). Gaussian components were used to reproduce the emission lines. 
Specifically, we used (1) a ‘systemic’ narrow (FWHM $<$ 700 km s$^{-1}$) Gaussian component for the [OIII] doublet, H$\beta$, H$\alpha$, [NII] and [SII] doublets  associated with unperturbed NLR emission; (2) a broad (FWHM$>$2000 km s$^{-1}$) Gaussian component to fit the BLR H$\beta$ and H$\alpha$ emission; (3) an ‘outflow’ (FWHM $>$ 700 km s$^{-1}$) Gaussian component for all forbidden and permitted emission lines to reproduce prominent and asymmetric wings associated with outflowing gas. One or more additional sets of Gaussian components were considered to reproduce more complex profiles, for instance, asymmetric BLR emission lines (see, for instance,  J1421+2241, Fig. \ref{fitsshen1}) and ‘extremely broad [OIII] profiles’ (for instance, J0843+0750, Fig. \ref{fitsshen2}; see also Fig. 11 in \citealt{Coatman2019}). 

For each set of Gaussian functions, we constrained the wavelength separation between emission lines within a given set of Gaussian profiles in accordance with atomic physics. This means that we constrained the velocity offset of the outflow Gaussians from the narrow (systemic) components to be the same for all the emission lines. Moreover, each emission line within a given set have the same FWHM. Finally, the relative flux of the two [N II]$\lambda\lambda$6548,5583 and [O III]$\lambda\lambda$4959,5007 components is fixed to 2.99 and the [S II] flux ratio is required to be within the range 0.44 $< f(\lambda 6716)/f(\lambda 6731) < $1.42; finally,  the H$\alpha$/H$\beta$ flux ratio has been constrained to be $> 2.85$ (\citealt{Osterbrock1989}).

The number of sets used to model the spectra was derived on the basis of the Bayesian information criterion (BIC; \citealt{Schwarz1978}), which uses differences in $\chi^2$ that penalise models with more free parameters (see e.g. \citealt{Harrison2016,Concas2019}).

\subsection{Modelling MgII, FeII, FeIII, H$\beta$ and [OIII] }\label{OIIImodel}

For the remaining targets at $z>$2.5, the H$\alpha$+[NII]+[SII] complex is not covered by available spectra. We therefore fitted H$\beta$+[OIII]+FeII systems together with the MgII+FeII+FeIII lines. Under the assumption that the H$\beta$ and MgII emission from BLR have the same kinematics (e.g. \citealt{Shen2012,Bisogni2017b}), as well that the optical and UV iron emission have the same width and velocity (e.g. \citealt{Bisogni2017b}), the simultaneous fit technique allows us to reduce, also for these spectra, the degeneracies between NLR, BLR and iron emission. We fitted the AGN continuum with two local power laws to the wavelength regions associated with the two systems (i.e., [$2600-3000$] \AA~ and [$4500-5500$] \AA). In addition to the optical FeII templates, we also considered one or two FeII+FeIII templates from \cite{Popovic2019} to model the iron emission underlying the MgII lines. The UV and optical iron templates are constrained to have the same widths and velocity shifts. Also in this case, we considered a combination of Gaussian sets to best reproduce the BLR and NLR emission (see above).    

We note that 1) the MgII$\lambda\lambda$2796,2803 doublet  is fitted with a single Gaussian, since the lines are unresolved in all spectra: MgII emission is generally dominated by the BLR emission, which width is much higher than the separation between the two member lines. Moreover, 2) the distinction between ’systemic’ and ‘outflow’ components in our best-fit results must be taken with caution, because of the degeneracies in the fit analysis, its dependency on the quality of the spectra (in terms of SNR and resolution), and the unknown intrinsic shape of ’systemic’ and ‘outflow’ emission line profiles (e.g. \citealt{Liu2013, Bae2016}).

\subsubsection{Main differences with respect to Shen (2016) and Coatman et al. (2019)}


Emission from the optical continuum and iron lines in \citet{Coatman2019} and \citet{Shen2016} are modelled with a local power-law plus a FeII template from \citet{Boroson1992}, using the wavelength regions just outside the H$\beta$-[OIII] complex. This model is then subtracted before fitting the BLR and NLR emission. In our analysis, all possible emitting contributions are fitted simultaneously, considering the entire wavelength range covered by FeII, H$\beta$ and [OIII] lines. Moreover, for the sources at z $\sim 2$, we used a unique power-law to model the continuum in the H$\beta$-[OIII] and H$\alpha$-[NII] regions, in order to be less affected by degeneracies between strong FeII and continuum emission. Finally, we preferred to use the FeII templates by \citet{Kovacevic2010}, which can allow a more robust separation between [OIII] and FeII emission at $\sim 5007 $\AA\ (see also e.g. Appendix A in \citealt{Perna2017a}).

In \citet{Shen2016} and \citet{Coatman2019} the [OIII] lines are fitted with two Gaussian components, one for the core and one for the prominent blueshifted wing. The only tied kinematics are those of the [OIII] and H$\beta$ core components.

\subsection{Best-fit analysis and results}

\begin{figure*}
\centering
\includegraphics[angle=90.,width=1.0\textwidth,height=0.7\textwidth]{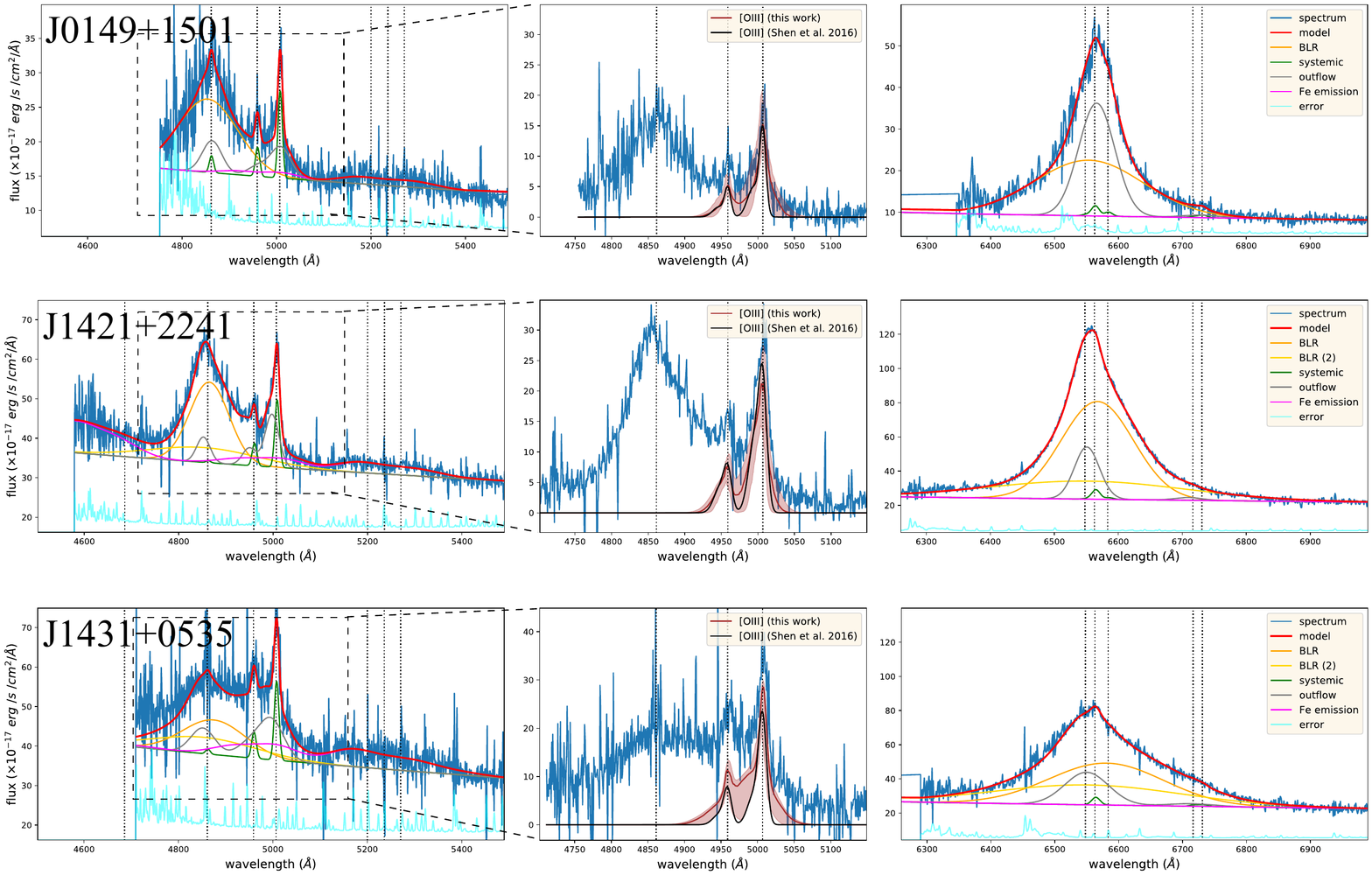}
\vskip-1cm
\includegraphics[angle=90.,width=1.0\textwidth,height=0.7\textwidth]{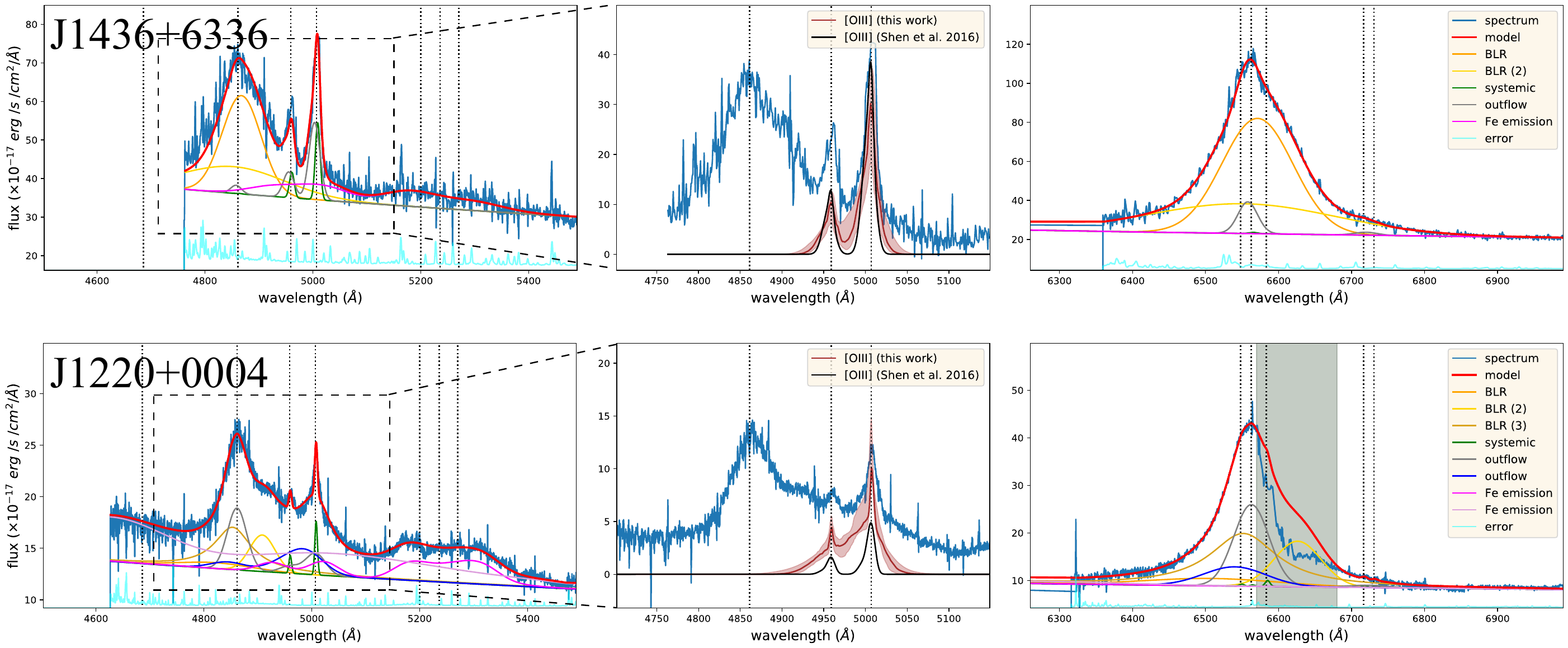}
\vskip-4cm
\caption{Parametrisation of the H$\beta$+[OIII]+FeII (left and middle panels) and H$\alpha$+[NII]+[SII] (right) regions of  $z\sim$2 QSO1 from the sample of \cite{Shen2016}. The red lines show the best-fit from multiple ’systemic’, ‘outflow and ‘broad’  Gaussian profiles to the line features, together with a single power-law to the continuum and one (two) template(s) to the FeII features (see text). Each curve with a different colour represents a distinct Gaussian set (or a FeII template) with same kinematic properties, as labeled in the figure. In the central panel we show, for each target, the spectrum around the H$\beta$+[OIII]  lines after the subtraction of the continuum. The red curve represents the best-fit retrieval model from Monte-Carlo (MC) fits, and the shaded area encompass 68\% of the MC sample; the black curve shows the [OIII] best-fit obtained by \cite{Shen2016}. }
\label{fitsshen1}
\end{figure*}

The spectral analysis was performed with a python routine using the cap-mpfit procedure (part of pPXF package by \citealt{Cappellari2017}) and performing Levenberg-Marquardt least-squares minimisation between the data and the model. The best-fit results are shown in Figs. \ref{fitsshen1}, \ref{fitsshen2}, \ref{fitsshen3} and reported in Tables \ref{shen} and \ref{OIIIbestfitparams}.

In order to estimate errors associated with our measurements, we used Monte Carlo simulations. For each modelled spectrum, we collected the fit results of 100 mock spectra obtained from the best-fit final models (red curves in Figs. \ref{fitsshen1}, \ref{fitsshen2} and \ref{fitsshen3}) and adding Gaussian random noise (taking as a reference the noise spectra shown with cyan curves in the figures). The errors, reported in Table \ref{shen}, have been derived by taking the range that contains 68.3\% of values evaluated from the obtained distributions for each Gaussian parameter/non-parametric measurement. Finally, we note that since line profiles generally are non-Gaussian and much broader than the spectral resolution, we do not correct the observed profiles for instrumental effects and report all values as measured. We also note that our estimated errors take into account (whenever possible) all systematic effects related to the (simultaneous) modelling of continuum, FeII, BLR and NLR emission.  In fact, our best-fit models are obtained, for each MC trial, starting from a random initialisation of the model parameters. This allows us to model the spectrum taking into account the possible degeneracies between the different emitting components.
As a consequence, the errors reported in Table \ref{shen} and \ref{OIIIbestfitparams} are up to a few orders of magnitude larger than those reported in \cite{Coatman2019}, more effectively reflecting the difficulties in such kinematic analysis.

For the sources for which we could fit both H$\alpha$ and H$\beta$, together with a single power-law from $\sim 4800$ to $\sim 7000$ \AA~ for the continuum, we can be quite confident about the contribution of BLR Balmer emission and the underlying continuum; all but J1220+0004 are also associated with weak FeII emission. These sources, therefore, provide the most robust kinematic measurements for the [OIII] gas. On the other hand, the sources at higher $z$ can be associated with higher systematic effects, because of the difficulties in modelling the local continuum underlying the H$\beta$ and MgII lines\footnote{For a proper modelling of the MgII+FeII+FeIII complex with a ‘global’ power-law continuum, a Balmer continuum emission as well as the doublet nature of the MgII should be taken into account (e.g. \citealt{Mejia2016,Varisco2018}).}, and the presence of strong iron emission (at least in the UV part). Overall, the differences in the analysis results between our work, \cite{Shen2016} and \cite{Coatman2019} highlight the difficulties in deriving robust [OIII] kinematic measurements for very luminous type 1 QSOs.

\begin{table}
\centering
\small
\begin{tabular}{lllll}\label{OIIIbestfitparams}
Object & Comp. & Amp. & $\Delta v$ & $FWHM$ \\
       &       &$10^{-17}$ cgs & km s$^{-1}$ & km s$^{-1}$   \\
\hline
J0149+1501 & 1st & $10.1_{-4.1}^{+3.4}$& $20_{-173}^{+138}$ & $273_{-153}^{+206}$\\
           & 2nd & $9.6_{-4.3}^{+3.0}$&$-63_{-283}^{+242}$&$1788_{-800}^{+1180}$\\
J1421+2241 & 1st & $13.9_{-5.6}^{+3.7}$&$22_{-200}^{+155}$&$337_{-183}^{+244}$\\
           & 2nd & $14.5_{-2.7}^{+3.8}$&$-380_{-158}^{+191}$&$1804_{-600}^{+508}$\\
J1431+0535 & 1st & $16.6_{-6.1}^{+3.4}$&$50_{-109}^{+105}$&$512_{-298}^{+187}$\\
           & 2nd & $12.1_{-2.1}^{+4.4}$&$-402_{-197}^{+244}$&$2845_{-1547}^{+854}$\\
J1436+6336 & 1st & $16.8_{-6.1}^{+3.2}$&$44_{-170}^{+110}$&$460_{-160}^{+170}$\\
           & 2nd & $16.1_{-7.3}^{+3.9}$&$-220_{-115}^{+150}$&$1870_{-550}^{+690}$\\
J1220+0004 & 1st & $5.5_{-1.5}^{+1.8}$&$28_{-43}^{+43}$&$252_{-53}^{+57}$\\
           & 2nd & $3.8_{-1.8}^{+3.0}$&$82_{-130}^{+131}$&$1625_{-640}^{+558}$\\
J0250-0757 & 1st & $8.0_{-4.9}^{+2.7}$&$-21_{-280}^{+121}$&$415_{-103}^{+165}$\\
           & 2nd & $4.8_{-1.9}^{3.2}$&$40_{-211}^{+113}$&$1442_{-646}^{+577}$\\
J0844+0503 & 1st & $8.2_{-2.1}^{3.6}$&$-85_{-57}^{+48}$&$1798_{-400}^{+300}$\\
           & 2nd & $9.1_{-1.3}^{+0.9}$&$-1737_{-359}^{+249}$&$3655_{-1047}^{+544}$\\
J0942+0422 & 1st & $14.5_{-4.1}^{+3.3}$&$45_{-125}^{+63}$&$399_{-91}^{+105}$\\
           & 2nd & $13.6_{-3.6}^{+2.4}$&$-164_{-199}^{+218}$&$1331_{-166}^{+232}$\\
           & 3rd & $3.5_{-1.1}^{+1.5}$&$44_{-189}^{+329}$&$3031_{-741}^{+1121}$\\
J0304-0008 & 1st & $58_{-6.9}^{+4.9}$&$102_{-92}^{+62}$&$415_{-74}^{+149}$\\
           & 2nd & $44.8_{-8.1}^{+9.8}$&$-54_{-42}^{+68}$&$788_{-93}^{+125}$\\
           & 3rd & $12.4_{-4.1}^{+3.3}$&$58_{-72}^{+47}$&$2165_{-395}^{+411}$\\
J0843+0750 & 1st & $13.1_{-1.8}^{+1.1}$&$60_{-8}^{+14}$&$411_{-39}^{+24}$\\
           & 2nd & $6.3_{-0.4}^{+0.3}$&$-34_{-27}^{+15}$&$1726_{-107}^{+86}$\\
J1019+0254 & 1st & $1.9_{-0.8}^{+0.9}$&$33_{-114}^{+95}$&$932_{-566}^{+420}$\\
           & 2nd & $2.0_{-0.7}^{+0.8}$&$78_{-102}^{+80}$&$2376_{-971}^{+866}$\\
J0259+0011 & 1st & $13.8_{3.4}^{+3.2}$&$14_{-92}^{+63}$&$441_{-148}^{+124}$\\
           & 2nd & $10.9_{-4.0}^{+3.9}$&$-10_{-154}^{+170}$&$1666_{-356}^{+317}$\\
J1034+0358 & 1st & $0.9_{-0.6}^{+3.54}$&$3_{-115}^{+107}$&$345_{-155}^{+150}$\\
           & 2nd & $1.6_{-0.4}^{+0.5}$&$-442_{-205}^{+190}$&$2555_{-720}^{+710}$\\
\hline 
\end{tabular}
  \caption{[OIII] best-fit parameters. For each component, we report the amplitude (in erg/s/cm$^2$/$\AA$), the centroid and the width of the best-fit Gaussian profile. }
\end{table}

Below we briefly comment of the best-fit results for each target. 

$\bullet$ J0149-1501: H$\beta$ and H$\alpha$  are redshifted at the edges of the H- and K-band respectively, where the filter transmission is very low. As a result, the bluest part of the line profiles cannot be modelled properly. The BLR emission is reproduced with one Gaussian component. Narrow emission can be well constrained only for the [OIII] doublet, which also present very broad line wings. A prominent ‘outflow’ component is required to reproduce the profiles of all emission lines (although the strong degeneracies do not allow us to well constrain this component for the Balmer lines). Our reconstructed [OIII] profiles present more pronounced blue and red wings with respect to those in \cite{Shen2016} and \cite{Coatman2019} , although all non-parametric velocities are consistent within 1$\sigma$.

$\bullet$ J1421+2241: the BLR is modelled with two Gaussian profiles; the NLR is fitted with a ’systemic’ and an ‘outflow’ component. Both blue and red wings of the [OIII] lines are blended with FeII and H$\beta$ emission. The [OIII] reconstructed profile is very similar to the ones in \cite{Shen2016} and \cite{Coatman2019}. 

$\bullet$ J1431+0535: the H$\beta$ line is severely affected by sky-line residuals; the broad H$\alpha$ from the BLR is strongly asymmetric. We modelled the BLR emission with two Gaussian components. Narrow emission can be well constrained only for the [OIII] doublet, which also present very broad line wings.  Our reconstructed OIII profiles present more pronounced blue and red wings with respect to those in \cite{Shen2016} and \cite{Coatman2019}, although the derived non-parametric velocities are consistent within 1$\sigma$ with theirs.

 $\bullet$ J1436+6336: the BLR lines are modelled with two Gaussian profiles; the NLR is fitted with a ’systemic’ and an ‘outflow’ component. Both blue and red wings of the [OIII] lines are mildly blended with FeII and H$\beta$. Our reconstructed [OIII] profiles present more pronounced blue and red wings with respect to those in  \cite{Shen2016} and \cite{Coatman2019}, although all non-parametric velocities are consistent within 2$\sigma$.

$\bullet$ J1220+0004: the H$\alpha$+[NII] complex is severely affected by telluric features; a significant portion of the profile is therefore masked during the spectral fitting. The BLR line profile is reconstructed on the basis of the observed H$\beta$ profile, with two BLR Gaussian sets and one Lorentzian profile. The NLR emission is fitted with a ’systemic’ component plus two very broad ‘outflow’ Gaussians. Two FeII templates are also required to reproduce the H$\beta$+[OIII]+FeII complex. Our reconstructed [OIII] profiles present more pronounced blue and red wings with respect to those in  \cite{Shen2016} and \cite{Coatman2019}. Under the assumption that H$\alpha$ and H$\beta$ have very similar profiles,  the flux excess around $\sim 5000$\AA~  cannot be associated with a H$\beta$ red wing, because of the absence of a similar feature in  H$\alpha$. On the other hand, FeII templates are not capable to reproduce the broad emission close to the [OIII] doublet. These arguments allow us to prefer a best-fit with very broad [OIII] lines. We note however that strong degeneracies could be still present in our fit, and that systematic errors could be significant for this target. 

$\bullet$ J2238-0921: this target shows very complex profiles in the H$\beta$+[OIII]+FeII and MgII+FeII+FeIII regions. We fit the MgII and H$\beta$ BLR emission with two Gaussian components. The iron emission is modelled with two templates. In the optical region, to reproduce the strong peak at $\sim 5250$\AA~ we include in the fit the FeIII and FeIV emission lines at 5270 and 5236\AA~ (but a strong degeneracy between these lines and FeII features is present at these wavelengths). No NLR emission associated with [OIII] is detected for this source, consistent with \cite[][see the detailed discussion in their paper]{Coatman2019} and at odds with \cite{Shen2016}. This source can be considered as a ‘weak [OIII]’ QSO1, with a CIV line blue-shift of $\sim$5500 km $^{-1}$ and REW$_{\rm CIV}\sim$ 10 \AA~ (see \citealt{Vietri2018}).

$\bullet$ J0250-0757: the BLR emission can be well reproduced with a single Gaussian profile;  [OIII] is modelled with two Gaussians; kinematic results are consistent with  \cite{Shen2016} and \cite{Coatman2019}.

$\bullet$ J0844-0503: the [OIII] profiles have similar characteristics to the core-ERQs. We modelled the H$\beta$ and MgII BLR emission with a single Gaussian; the NRL emission is fitted with two very broad ‘outflow’ Gaussians.  Our reconstructed [OIII] profiles present more pronounced blue and red wings with respect to those in \cite{Shen2016} and \cite{Coatman2019}.

 $\bullet$ J0942-0422: the BLR emission can be well reproduced with a single Gaussian profile; the [OIII] is modelled with three Gaussians.  Our reconstructed [OIII] profiles present more pronounced red wings with respect to those in  \cite{Shen2016} and \cite{Coatman2019}.

$\bullet$ J0304-0008: the BLR emission can be well reproduced with a single Gaussian profile; the [OIII] is modelled with three Gaussians.  Our reconstructed [OIII] profiles present more pronounced wings with respect to those in \cite{Shen2016}, consistent with  \cite{Coatman2019}.

 $\bullet$ J0843+0750:  the BLR emission can be well reproduced with a single Gaussian profile; the [OIII] is modelled with two Gaussians.  Our reconstructed [OIII] profiles present more pronounced red wings with respect to those in  \cite{Shen2016}, consistent with  \cite{Coatman2019}.

$\bullet$ J1019+0254: the BLR emission can be well reproduced with a single Gaussian profile; the [OIII] is modelled with two Gaussians.  Our reconstructed [OIII] profiles present more pronounced wings with respect to those in  \cite{Shen2016}, consistent with  \cite{Coatman2019}

$\bullet$ J0259+0011: the BLR emission can be well reproduced with a single Gaussian profile; the [OIII] is modelled with two Gaussians.  Our reconstructed [OIII] profiles present more pronounced wings with respect to those in  \cite{Shen2016}, consistent with  \cite{Coatman2019}

$\bullet$ J1034+0358:  the BLR emission can be well reproduced with a single Gaussian profile; the [OIII] is modelled with two Gaussians.  Our reconstructed [OIII] profiles present more pronounced wings with respect to those in  \cite{Shen2016} and \cite{Coatman2019}.

\begin{figure*}
\centering
\vskip-1cm
\includegraphics[angle=90.,width=1.0\textwidth,height=0.7\textwidth]{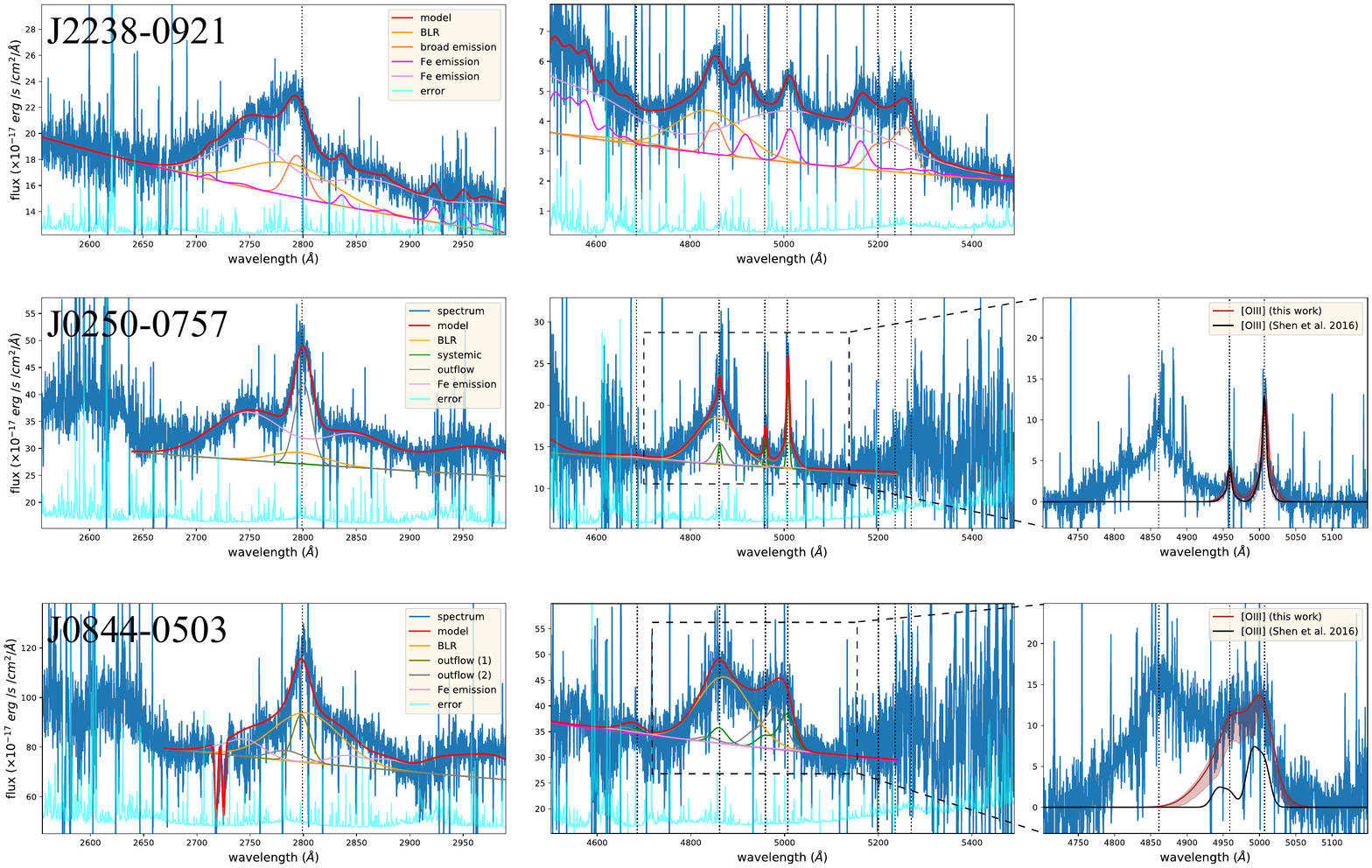}
\vskip-5cm
\includegraphics[angle=90.,width=1.0\textwidth,height=0.7\textwidth]{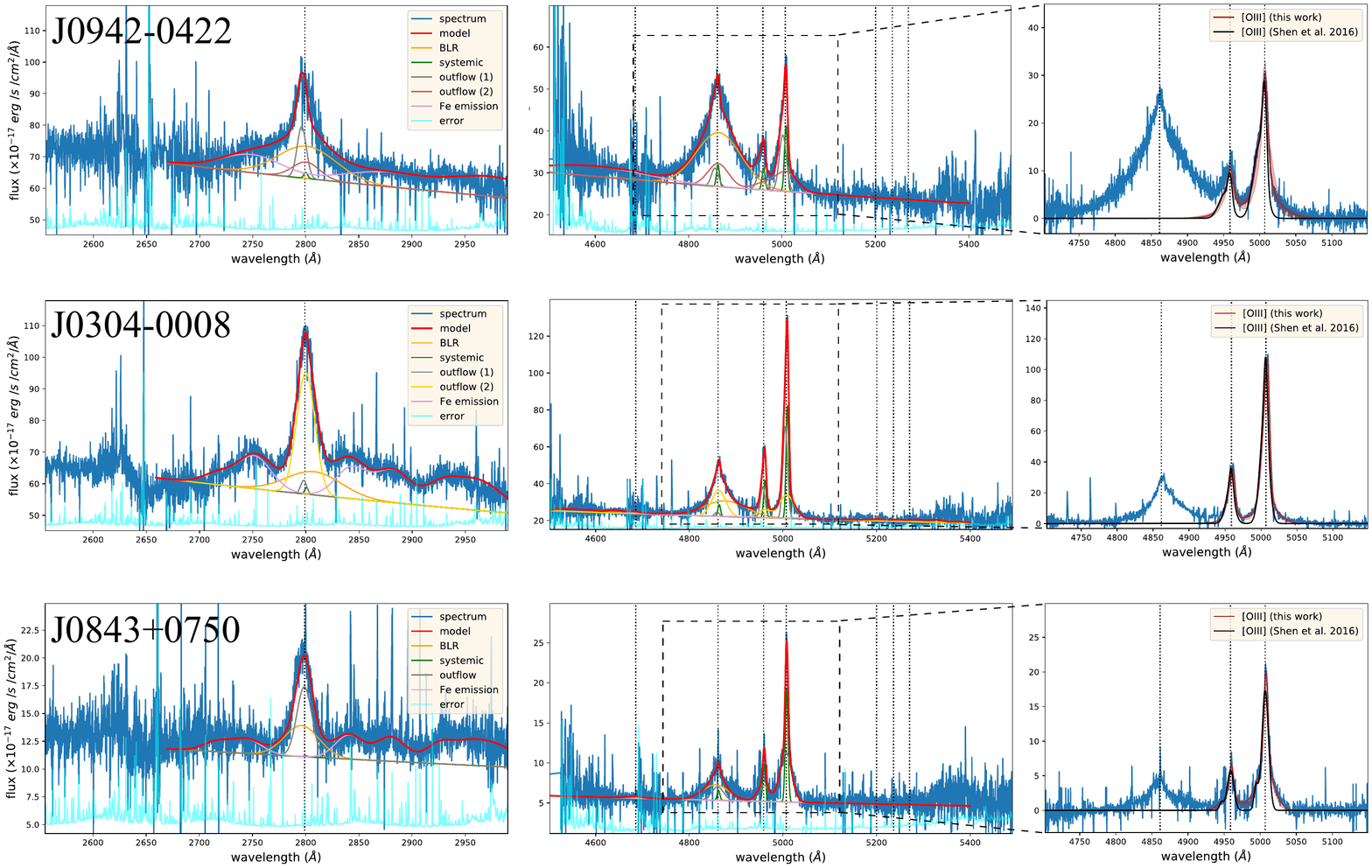}
\vskip-0.5cm
\caption{Parametrisation of the MgII+FeII+FeIII (left) and H$\beta$+[OIII]+FeII (middle and right panels) )  of $z>2.5$ QSO1 from the sample of \cite{Shen2016}. The red lines show the best-fit from multiple ’systemic’, ‘outflow and ‘broad’  Gaussian profiles to the line features, together with a two local power-laws to the continuum and one (two) template(s) to the FeII features (see text). Each curve with a different colour represent a distinct Gaussian set (or a FeII template) with same kinematic properties, as labeled in the figure. In the right panel we show, for each target, the spectrum around the H$\beta$+[OIII] lines after the subtraction of the continuum. The red curve represents the best-fit retrieval model from MC fits, and the shaded area encompass 68\% of the MC sample; the black curve shows the [OIII] best-fit obtained by \cite{Shen2016}. For J2238-0921, we do not detect any [OIII] emission line, consistent with \cite{Coatman2019} and at odds with \cite{Shen2016}:  the peaks at $\sim 4900$, $5000$ and $5150$\AA~ are associated with FeII emission (precisely, the $^6S$ iron line group, according to the ordering in \citealt{Kovacevic2010}).}
\label{fitsshen2}
\end{figure*}

\begin{figure*}
\centering
\vskip-1cm
\includegraphics[angle=90.,width=1.0\textwidth,height=0.7\textwidth]{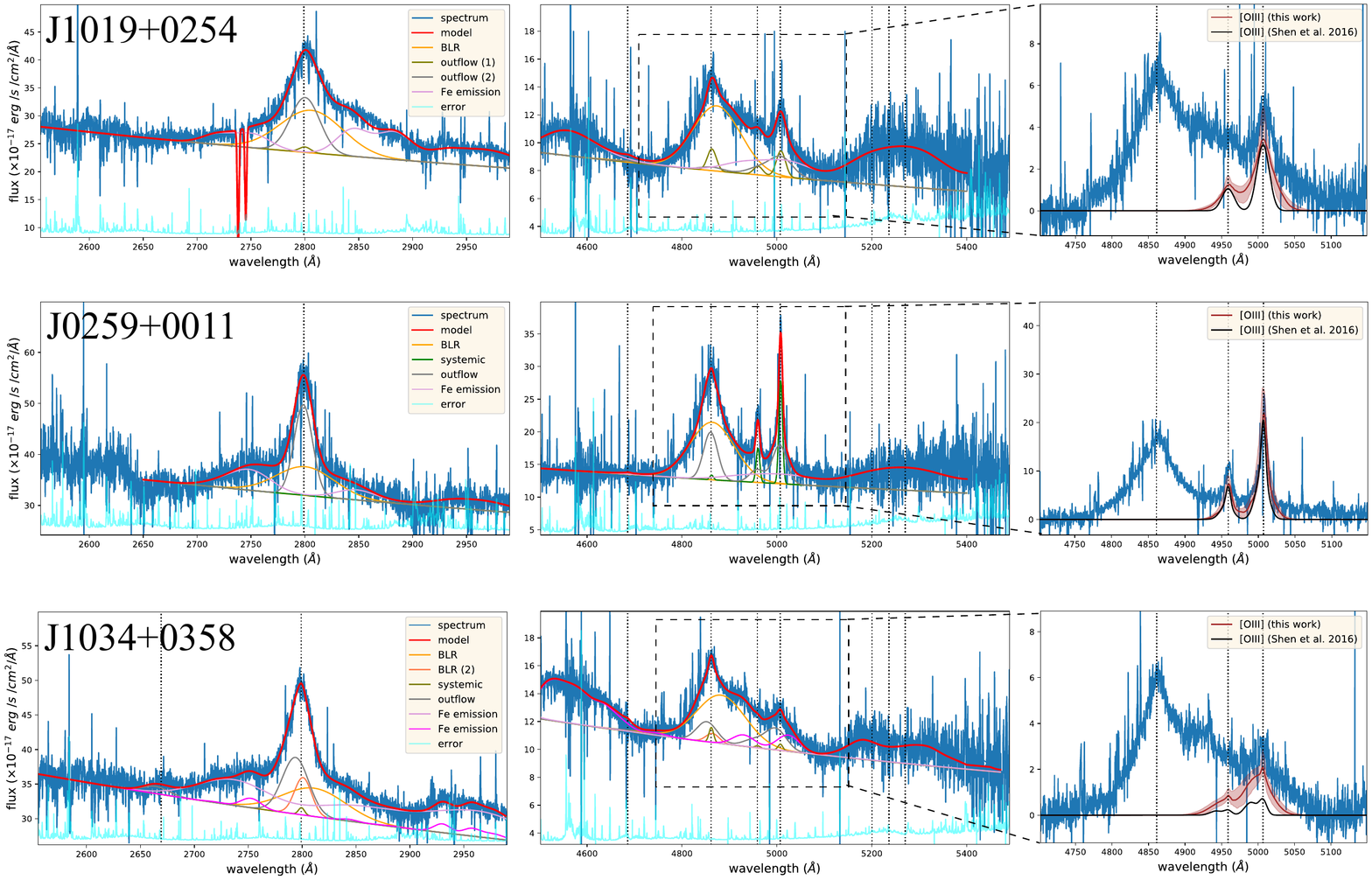}
\caption{Continuation of Fig. 6}
\label{fitsshen3}
\end{figure*}

\end{appendix}

\end{document}